\documentclass[nojss]{jss}

\pdfoutput=1 
\usepackage{amsmath, amsfonts, amssymb, amsthm, graphicx}
\usepackage{bm}
\usepackage{url}

\usepackage{algorithm}
\usepackage{algorithmic}
\usepackage{subcaption}
\usepackage{multirow}
\usepackage{stmaryrd}
\usepackage{tikz}
\usepackage{tkz-graph}
\tikzstyle{vertex}=[circle,fill=white,draw]
\usetikzlibrary{arrows.meta,calc,decorations.markings,math,arrows.meta}

\newcommand{\fct}[1]{\code{#1()}}

\floatstyle{ruled}
\newfloat{algorithm}{tbp}{loa}
\providecommand{\algorithmname}{Algorithm}
\floatname{algorithm}{\protect\algorithmname}

\author{Yuqing Pan\\Florida State University
	\And Qing Mai\\Florida State University
	\And Xin Zhang\\Florida State University}
\Plainauthor{Yuqing Pan, Qing Mai, Xin Zhang}

\title{TULIP: A Toolbox for Linear Discriminant Analysis with Penalties\thanks{\small Research for this paper was supported in part by grants CCF-1617691 and DMS-1613154 from the U.S. National Science Foundation.  }}
\Shorttitle{TULIP in \proglang{R}}

\Abstract{
	Linear discriminant analysis (LDA) is a powerful tool in building classifiers with easy computation and interpretation. Recent advancements in science technology have led to the popularity of datasets with high dimensions, high orders and complicated structure. Such datasetes motivate the generalization of LDA in various research directions. The $\tt R$ package $\tt TULIP$ integrates several popular high-dimensional LDA-based methods and provides a comprehensive and user-friendly toolbox for linear, semi-parametric and tensor-variate classification. Functions are included for model fitting, cross validation and prediction. In addition, motivated by datasets with diverse sources of predictors, we further include functions for covariate adjustment. Our package is carefully tailored for low storage and high computation efficiency. Moreover, our package is the first $\tt R$ package for many of these methods, providing great convenience to researchers in this area.}

\Keywords{Discriminant analysis, penalty, multicategory classification, sparsity, tensor data}

\Address{
	Yuqing Pan, Qing Mai, Xin Zhang\\
	Department of Statistics\\
	Florida State University\\
	Tallahassee, FL 32306, United States of America\\
	E-mail: \email{yuqing.pan@stat.fsu.edu}, \email{mai@stat.fsu.edu}, \email{henry@stat.fsu.edu}\\
	URL: \url{https://ani.stat.fsu.edu/~mai/}, \url{https://ani.stat.fsu.edu/~henry/}
}

\begin{document}
	
	\global\long\def\mba{\mathbf{a}}
	\global\long\def\mbA{\mathbf{A}}
	\global\long\def\mbb{\mathbf{b}}
	\global\long\def\mbB{\mathbf{B}}
	\global\long\def\mbc{\mathbf{c}}
	\global\long\def\mbC{\mathbf{C}}
	\global\long\def\mbd{\mathbf{d}}
	\global\long\def\mbD{\mathbf{D}}
	\global\long\def\mbe{\mathbf{e}}
	\global\long\def\mbE{\mathbf{E}}
	\global\long\def\mbf{\mathbf{f}}
	\global\long\def\mbF{\mathbf{F}}
	\global\long\def\mbg{\mathbf{g}}
	\global\long\def\mbG{\mathbf{G}}
	\global\long\def\mbh{\mathbf{h}}
	\global\long\def\mbH{\mathbf{H}}
	\global\long\def\mbi{\mathbf{i}}
	\global\long\def\mbI{\mathbf{I}}
	\global\long\def\mbj{\mathbf{j}}
	\global\long\def\mbJ{\mathbf{J}}
	\global\long\def\mbk{\mathbf{k}}
	\global\long\def\mbK{\mathbf{K}}
	\global\long\def\mbl{\mathbf{l}}
	\global\long\def\mbL{\mathbf{L}}
	\global\long\def\mbm{\mathbf{m}}
	\global\long\def\mbM{\mathbf{M}}
	\global\long\def\mbn{\mathbf{n}}
	\global\long\def\mbN{\mathbf{N}}
	\global\long\def\mbo{\mathbf{o}}
	\global\long\def\mbO{\mathbf{O}}
	\global\long\def\mbp{\mathbf{p}}
	\global\long\def\mbP{\mathbf{P}}
	\global\long\def\mbq{\mathbf{q}}
	\global\long\def\mbQ{\mathbf{Q}}
	\global\long\def\mbr{\mathbf{r}}
	\global\long\def\mbR{\mathbf{R}}
	\global\long\def\mbs{\mathbf{s}}
	\global\long\def\mbS{\mathbf{S}}
	\global\long\def\mbt{\mathbf{t}}
	\global\long\def\mbT{\mathbf{T}}
	\global\long\def\mbu{\mathbf{u}}
	\global\long\def\mbU{\mathbf{U}}
	\global\long\def\mbv{\mathbf{v}}
	\global\long\def\mbV{\mathbf{V}}
	\global\long\def\mbw{\mathbf{w}}
	\global\long\def\mbW{\mathbf{W}}
	\global\long\def\mbx{\mathbf{x}}
	\global\long\def\mbX{\mathbf{X}}
	\global\long\def\mby{\mathbf{y}}
	\global\long\def\mbY{\mathbf{Y}}
	\global\long\def\mbz{\mathbf{z}}
	\global\long\def\mbZ{\mathbf{Z}}
	\global\long\def\barmbZ{\bar{\mathbf{Z}}}
	\global\long\def\hatmba{\widehat{\mathbf{a}}}
	\global\long\def\hatmbA{\widehat{\mathbf{A}}}
	\global\long\def\hatmbb{\widehat{\mathbf{b}}}
	\global\long\def\hatmbB{\widehat{\mathbf{B}}}
	\global\long\def\hatmbc{\widehat{\mathbf{c}}}
	\global\long\def\hatmbC{\widehat{\mathbf{C}}}
	\global\long\def\hatmbd{\widehat{\mathbf{d}}}
	\global\long\def\hatmbD{\widehat{\mathbf{D}}}
	\global\long\def\hatmbe{\widehat{\mathbf{e}}}
	\global\long\def\hatmbE{\widehat{\mathbf{E}}}
	\global\long\def\hatmbf{\widehat{\mathbf{f}}}
	\global\long\def\hatmbF{\widehat{\mathbf{F}}}
	\global\long\def\hatmbg{\widehat{\mathbf{g}}}
	\global\long\def\hatmbG{\widehat{\mathbf{G}}}
	\global\long\def\hatmbh{\widehat{\mathbf{h}}}
	\global\long\def\hatmbH{\widehat{\mathbf{H}}}
	\global\long\def\hatmbi{\widehat{\mathbf{i}}}
	\global\long\def\hatmbI{\widehat{\mathbf{I}}}
	\global\long\def\hatmbj{\widehat{\mathbf{j}}}
	\global\long\def\hatmbJ{\widehat{\mathbf{J}}}
	\global\long\def\hatmbk{\widehat{\mathbf{k}}}
	\global\long\def\hatmbK{\widehat{\mathbf{K}}}
	\global\long\def\hatmbl{\widehat{\mathbf{l}}}
	\global\long\def\hatmbL{\widehat{\mathbf{L}}}
	\global\long\def\hatmbm{\widehat{\mathbf{m}}}
	\global\long\def\hatmbM{\widehat{\mathbf{M}}}
	\global\long\def\hatmbn{\widehat{\mathbf{n}}}
	\global\long\def\hatmbN{\widehat{\mathbf{N}}}
	\global\long\def\hatmbo{\widehat{\mathbf{o}}}
	\global\long\def\hatmbO{\widehat{\mathbf{O}}}
	\global\long\def\hatmbp{\widehat{\mathbf{p}}}
	\global\long\def\hatmbP{\widehat{\mathbf{P}}}
	\global\long\def\hatmbq{\widehat{\mathbf{q}}}
	\global\long\def\hatmbQ{\widehat{\mathbf{Q}}}
	\global\long\def\hatmbr{\widehat{\mathbf{r}}}
	\global\long\def\hatmbR{\widehat{\mathbf{R}}}
	\global\long\def\hatmbs{\widehat{\mathbf{s}}}
	\global\long\def\hatmbS{\widehat{\mathbf{S}}}
	\global\long\def\hatmbt{\widehat{\mathbf{t}}}
	\global\long\def\hatmbT{\widehat{\mathbf{T}}}
	\global\long\def\hatmbu{\widehat{\mathbf{u}}}
	\global\long\def\hatmbU{\widehat{\mathbf{U}}}
	\global\long\def\hatmbv{\widehat{\mathbf{v}}}
	\global\long\def\hatmbV{\widehat{\mathbf{V}}}
	\global\long\def\hatmbw{\widehat{\mathbf{w}}}
	\global\long\def\hatmbW{\widehat{\mathbf{W}}}
	\global\long\def\hatmbx{\widehat{\mathbf{x}}}
	\global\long\def\hatmbX{\widehat{\mathbf{X}}}
	\global\long\def\hatmby{\widehat{\mathbf{y}}}
	\global\long\def\hatmbY{\widehat{\mathbf{Y}}}
	\global\long\def\hatmbz{\widehat{\mathbf{z}}}
	\global\long\def\hatmbZ{\widehat{\mathbf{Z}}}

	\global\long\def\tilmba{\widetilde{\mathbf{a}}}
	\global\long\def\tilmbA{\widetilde{\mathbf{A}}}
	\global\long\def\tilmbb{\widetilde{\mathbf{b}}}
	\global\long\def\tilmbB{\widetilde{\mathbf{B}}}
	\global\long\def\tilmbc{\widetilde{\mathbf{c}}}
	\global\long\def\tilmbC{\widetilde{\mathbf{C}}}
	\global\long\def\tilmbd{\widetilde{\mathbf{d}}}
	\global\long\def\tilmbD{\widetilde{\mathbf{D}}}
	\global\long\def\tilmbe{\widetilde{\mathbf{e}}}
	\global\long\def\tilmbE{\widetilde{\mathbf{E}}}
	\global\long\def\tilmbf{\widetilde{\mathbf{f}}}
	\global\long\def\tilmbF{\widetilde{\mathbf{F}}}
	\global\long\def\tilmbg{\widetilde{\mathbf{g}}}
	\global\long\def\tilmbG{\widetilde{\mathbf{G}}}
	\global\long\def\tilmbh{\widetilde{\mathbf{h}}}
	\global\long\def\tilmbH{\widetilde{\mathbf{H}}}
	\global\long\def\tilmbi{\widetilde{\mathbf{i}}}
	\global\long\def\tilmbI{\widetilde{\mathbf{I}}}
	\global\long\def\tilmbj{\widetilde{\mathbf{j}}}
	\global\long\def\tilmbJ{\widetilde{\mathbf{J}}}
	\global\long\def\tilmbk{\widetilde{\mathbf{k}}}
	\global\long\def\tilmbK{\widetilde{\mathbf{K}}}
	\global\long\def\tilmbl{\widetilde{\mathbf{l}}}
	\global\long\def\tilmbL{\widetilde{\mathbf{L}}}
	\global\long\def\tilmbm{\widetilde{\mathbf{m}}}
	\global\long\def\tilmbM{\widetilde{\mathbf{M}}}
	\global\long\def\tilmbn{\widetilde{\mathbf{n}}}
	\global\long\def\tilmbN{\widetilde{\mathbf{N}}}
	\global\long\def\tilmbo{\widetilde{\mathbf{o}}}
	\global\long\def\tilmbO{\widetilde{\mathbf{O}}}
	\global\long\def\tilmbp{\widetilde{\mathbf{p}}}
	\global\long\def\tilmbP{\widetilde{\mathbf{P}}}
	\global\long\def\tilmbq{\widetilde{\mathbf{q}}}
	\global\long\def\tilmbQ{\widetilde{\mathbf{Q}}}
	\global\long\def\tilmbr{\widetilde{\mathbf{r}}}
	\global\long\def\tilmbR{\widetilde{\mathbf{R}}}
	\global\long\def\tilmbs{\widetilde{\mathbf{s}}}
	\global\long\def\tilmbS{\widetilde{\mathbf{S}}}
	\global\long\def\tilmbt{\widetilde{\mathbf{t}}}
	\global\long\def\tilmbT{\widetilde{\mathbf{T}}}
	\global\long\def\tilmbu{\widetilde{\mathbf{u}}}
	\global\long\def\tilmbU{\widetilde{\mathbf{U}}}
	\global\long\def\tilmbv{\widetilde{\mathbf{v}}}
	\global\long\def\tilmbV{\widetilde{\mathbf{V}}}
	\global\long\def\tilmbw{\widetilde{\mathbf{w}}}
	\global\long\def\tilmbW{\widetilde{\mathbf{W}}}
	\global\long\def\tilmbx{\widetilde{\mathbf{x}}}
	\global\long\def\tilmbX{\widetilde{\mathbf{X}}}
	\global\long\def\tilmby{\widetilde{\mathbf{y}}}
	\global\long\def\tilmbY{\widetilde{\mathbf{Y}}}
	\global\long\def\tilmbz{\widetilde{\mathbf{z}}}
	\global\long\def\tilmbZ{\widetilde{\mathbf{Z}}}
	
	\renewcommand{\Vec}{\mathrm{Vec}}
	\newcommand{\bSigma}{{\bm \Sigma}}
	\newcommand{\bzero}{\mathbf{0}}
	
	\newcommand{\bX}{\mathbf{X}}
	\newcommand{\bY}{\mathbf{Y}}
	\newcommand{\bS}{\mathbf{S}}
	\newcommand{\bx}{\mathbf{x}}
	\newcommand{\bV}{\mathbf{V}}
	\newcommand{\bv}{\mathbf{v}}
	\newcommand{\bG}{\mathbf{G}}
	\newcommand{\bbeta}{{\bm\beta}}
	\newcommand{\bgamma}{{\bm\gamma}}
	\newcommand{\bGamma}{{\bm\Gamma}}
	\newcommand{\bLambda}{{\bm\Lambda}}
	
	\newcommand{\bA}{\mathbf{A}}
	\newcommand{\bB}{\mathbf{B}}
	\newcommand{\bC}{\mathbf{C}}
	\newcommand{\bD}{\mathbf{D}}
	\newcommand{\bI}{\mathbf{I}}
	\newcommand{\bJ}{\mathbf{J}}
	\newcommand{\bR}{\mathbf{R}}
	\newcommand{\bu}{\mathbf{u}}
	\newcommand{\bz}{\mathbf{z}}
	\newcommand{\f}{\mathbf{f}}
	\newcommand{\bzeta}{{\bm \zeta}}
	\newcommand{\btheta}{{\bm \theta}}
	\newcommand{\bP}{\mathbf{P}}
	\newcommand{\bT}{\mathbf{T}}
	\newcommand{\bZ}{\mathbf{Z}}
	\newcommand{\bepsilon}{\bm \epsilon}
	\newcommand{\bmu}{\bm \mu}
	\newcommand{\bxi}{\bm \xi}
	\newcommand{\bW}{\mathbf{W}}
	\newcommand{\bOmega}{\bm \Omega}
	\newcommand{\bU}{\mathbf{U}}
	\newcommand{\bTheta}{\bm \Theta}
	\newcommand{\bpi}{\bm \pi}
	\newcommand{\balpha}{\bm \alpha}
	\newcommand{\bdelta}{\bm \delta}
	\newcommand{\bF}{\mathbf{F}}

	\global\long\def\hatf{\widehat{f}}
	
	\global\long\def\bolalpha{\boldsymbol{\alpha}}
	\global\long\def\bolbeta{\boldsymbol{\beta}}
	\global\long\def\bolgamma{\boldsymbol{\gamma}}
	\global\long\def\boldelta{\boldsymbol{\delta}}
	\global\long\def\bolepsilon{\boldsymbol{\epsilon}}
	\global\long\def\bolzeta{\boldsymbol{\zeta}}
	\global\long\def\boleta{\boldsymbol{\eta}}
	\global\long\def\boltheta{\boldsymbol{\theta}}
	\global\long\def\bolkappa{\boldsymbol{\kappa}}
	\global\long\def\bollambda{\boldsymbol{\lambda}}
	\global\long\def\bolmu{\boldsymbol{\mu}}
	\global\long\def\bolnu{\boldsymbol{\nu}}
	\global\long\def\bolxi{\boldsymbol{\xi}}
	\global\long\def\bolpi{\boldsymbol{\pi}}
	\global\long\def\bolrho{\boldsymbol{\rho}}
	\global\long\def\bolsigma{\boldsymbol{\sigma}}
	\global\long\def\boltau{\boldsymbol{\tau}}
	\global\long\def\bolphi{\boldsymbol{\phi}}
	\global\long\def\bolchi{\boldsymbol{\chi}}
	\global\long\def\bolpsi{\boldsymbol{\psi}}
	\global\long\def\bolomega{\boldsymbol{\omega}}
	\global\long\def\bolGamma{\boldsymbol{\Gamma}}
	\global\long\def\bolDelta{\boldsymbol{\Delta}}
	\global\long\def\bolTheta{\boldsymbol{\Theta}}
	\global\long\def\bolLambda{\boldsymbol{\Lambda}}
	\global\long\def\bolPi{\boldsymbol{\Pi}}
	\global\long\def\bolSigma{\boldsymbol{\Sigma}}
	\global\long\def\bolPhi{\boldsymbol{\Phi}}
	\global\long\def\bolPsi{\boldsymbol{\Psi}}
	\global\long\def\bolOmega{\boldsymbol{\Omega}}

	\global\long\def\hatbolalpha{\widehat{\boldsymbol{\alpha}}}
	\global\long\def\hatbolbeta{\widehat{\boldsymbol{\beta}}}
	\global\long\def\hatbolgamma{\widehat{\boldsymbol{\gamma}}}
	\global\long\def\hatboldelta{\widehat{\boldsymbol{\delta}}}
	\global\long\def\hatbolepsilon{\widehat{\boldsymbol{\epsilon}}}
	\global\long\def\hatbolzeta{\widehat{\boldsymbol{\zeta}}}
	\global\long\def\hatboleta{\widehat{\boldsymbol{\eta}}}
	\global\long\def\hatboltheta{\widehat{\boldsymbol{\theta}}}
	\global\long\def\hatbolkappa{\widehat{\boldsymbol{\kappa}}}
	\global\long\def\hatbollambda{\widehat{\boldsymbol{\lambda}}}
	\global\long\def\hatbolmu{\widehat{\boldsymbol{\mu}}}
	\global\long\def\hatbolnu{\widehat{\boldsymbol{\nu}}}
	\global\long\def\hatbolxi{\widehat{\boldsymbol{\xi}}}
	\global\long\def\hatbolpi{\widehat{\boldsymbol{\pi}}}
	\global\long\def\hatbolrho{\widehat{\boldsymbol{\rho}}}
	\global\long\def\hatbolsigma{\widehat{\boldsymbol{\sigma}}}
	\global\long\def\hatboltau{\widehat{\boldsymbol{\tau}}}
	\global\long\def\hatbolphi{\widehat{\boldsymbol{\phi}}}
	\global\long\def\hatbolchi{\widehat{\boldsymbol{\chi}}}
	\global\long\def\hatbolpsi{\widehat{\boldsymbol{\psi}}}
	\global\long\def\hatbolomega{\widehat{\boldsymbol{\omega}}}
	\global\long\def\hatbolGamma{\widehat{\boldsymbol{\Gamma}}}
	\global\long\def\hatbolDelta{\widehat{\boldsymbol{\Delta}}}
	\global\long\def\hatbolTheta{\widehat{\boldsymbol{\Theta}}}
	\global\long\def\hatbolLambda{\widehat{\boldsymbol{\Lambda}}}
	\global\long\def\hatbolPi{\widehat{\boldsymbol{\Pi}}}
	\global\long\def\hatbolSigma{\widehat{\boldsymbol{\Sigma}}}
	\global\long\def\hatbolPhi{\widehat{\boldsymbol{\Phi}}}
	\global\long\def\hatbolPsi{\widehat{\boldsymbol{\Psi}}}
	\global\long\def\hatbolOmega{\widehat{\boldsymbol{\Omega}}}

	\global\long\def\tilbolalpha{\widetilde{\boldsymbol{\alpha}}}
	\global\long\def\tilbolbeta{\widetilde{\boldsymbol{\beta}}}
	\global\long\def\tilbolgamma{\widetilde{\boldsymbol{\gamma}}}
	\global\long\def\tilboldelta{\widetilde{\boldsymbol{\delta}}}
	\global\long\def\tilbolepsilon{\widetilde{\boldsymbol{\epsilon}}}
	\global\long\def\tilbolzeta{\widetilde{\boldsymbol{\zeta}}}
	\global\long\def\tilboleta{\widetilde{\boldsymbol{\eta}}}
	\global\long\def\tilboltheta{\widetilde{\boldsymbol{\theta}}}
	\global\long\def\tilbolkappa{\widetilde{\boldsymbol{\kappa}}}
	\global\long\def\tilbollambda{\widetilde{\boldsymbol{\lambda}}}
	\global\long\def\tilbolmu{\widetilde{\boldsymbol{\mu}}}
	\global\long\def\tilbolnu{\widetilde{\boldsymbol{\nu}}}
	\global\long\def\tilbolxi{\widetilde{\boldsymbol{\xi}}}
	\global\long\def\tilbolpi{\widetilde{\boldsymbol{\pi}}}
	\global\long\def\tilbolrho{\widetilde{\boldsymbol{\rho}}}
	\global\long\def\tilbolsigma{\widetilde{\boldsymbol{\sigma}}}
	\global\long\def\tilboltau{\widetilde{\boldsymbol{\tau}}}
	\global\long\def\tilbolphi{\widetilde{\boldsymbol{\phi}}}
	\global\long\def\tilbolchi{\widetilde{\boldsymbol{\chi}}}
	\global\long\def\tilbolpsi{\widetilde{\boldsymbol{\psi}}}
	\global\long\def\tilbolomega{\widetilde{\boldsymbol{\omega}}}
	\global\long\def\tilbolGamma{\widetilde{\boldsymbol{\Gamma}}}
	\global\long\def\tilbolDelta{\widetilde{\boldsymbol{\Delta}}}
	\global\long\def\tilbolTheta{\widetilde{\boldsymbol{\Theta}}}
	\global\long\def\tilbolLambda{\widetilde{\boldsymbol{\Lambda}}}
	\global\long\def\tilbolPi{\widetilde{\boldsymbol{\Pi}}}
	\global\long\def\tilbolSigma{\widetilde{\boldsymbol{\Sigma}}}
	\global\long\def\tilbolPhi{\widetilde{\boldsymbol{\Phi}}}
	\global\long\def\tilbolPsi{\widetilde{\boldsymbol{\Psi}}}
	\global\long\def\tilbolOmega{\widetilde{\boldsymbol{\Omega}}}

	\global\long\def\barbolmu{\overline{\bolmu}}
	\global\long\def\barmbX{\overline{\mbX}}

	\global\long\def\mbbR{\mathbb{R}}
	\global\long\def\mbbS{\mathbb{S}}
	\global\long\def\mbbX{\mathbb{X}}
	\global\long\def\mbbY{\mathbb{Y}}
	\global\long\def\mbbZ{\mathbb{Z}}
	\global\long\def\mbbU{\mathbb{U}}
	
	\global\long\def\calA{\mathcal{A}}
	\global\long\def\calB{\mathcal{B}}
	\global\long\def\calC{\mathcal{C}}
	\global\long\def\calD{\mathcal{D}}
	\global\long\def\calE{\mathcal{E}}
	\global\long\def\calF{\mathcal{F}}
	\global\long\def\calG{\mathcal{G}}
	\global\long\def\calH{\mathcal{H}}
	\global\long\def\calI{\mathcal{I}}
	\global\long\def\calJ{\mathcal{J}}
	\global\long\def\calK{\mathcal{K}}
	\global\long\def\calL{\mathcal{L}}
	\global\long\def\calM{\mathcal{M}}
	\global\long\def\calN{\mathcal{N}}
	\global\long\def\calO{\mathcal{O}}
	\global\long\def\calP{\mathcal{P}}
	\global\long\def\calQ{\mathcal{Q}}
	\global\long\def\calR{\mathcal{R}}
	\global\long\def\calS{\mathcal{S}}
	\global\long\def\calT{\mathcal{T}}
	\global\long\def\calU{\mathcal{U}}
	\global\long\def\calV{\mathcal{V}}
	\global\long\def\calW{\mathcal{W}}

	\global\long\def\mbell{\boldsymbol{\ell}}
	\global\long\def\bolell{\boldsymbol{\ell}}
	\global\long\def\mbzero{\mathbf{0}}

	\global\long\def\bolPhio{\boldsymbol{\Phi}_{0}}
	\global\long\def\bolOmegao{\boldsymbol{\Omega}_{0}}

	\global\long\def\bolSigmaX{\bolSigma_{\mbX}}
	\global\long\def\bolSigmaY{\bolSigma_{\mbY}}
	\global\long\def\bolSigmaXY{\boldsymbol{\Sigma}_{\mbX\mbY}}
	\global\long\def\mbSX{\mathbf{S}_{\mbX}}
	\global\long\def\mbSY{\mathbf{S}_{\mbY}}
	\global\long\def\mbSXY{\mathbf{S}_{\mbX\mbY}}
	\global\long\def\mbSYX{\mathbf{S}_{\mbY\mbX}}
	\global\long\def\mbRYX{\mathbf{S}_{\mbY|\mbX}}
	\global\long\def\mbRXY{\mathbf{S}_{\mbX|\mbY}}
	\global\long\def\mbSc{\mbS_{\mbC}}
	\global\long\def\mbSd{\mbS_{\mbD}}

	\global\long\def\sumn{\sum_{i=1}^{n}}

	\global\long\def\E{\mathrm{E}}
	\global\long\def\F{\mathrm{F}}
	\global\long\def\J{\mathrm{J}}
	\global\long\def\H{\mathrm{H}}
	\global\long\def\G{\mathrm{G}}
	\global\long\def\Cov{\mathrm{cov}}
	\global\long\def\Corr{\mathrm{corr}}
	\global\long\def\Var{\mathrm{var}}
	\global\long\def\dimension{\mathrm{dim}}
	\global\long\def\spn{\mathrm{span}}
	\global\long\def\vech{\mathrm{vech}}
	\global\long\def\vecc{\mathrm{vec}}
	\global\long\def\Prob{\mathrm{Pr}}
	\global\long\def\Env{\mathrm{env}}
	\global\long\def\tr{\mathrm{tr}}
	\global\long\def\dg{\mathrm{diag}}
	\global\long\def\asyVar{\mathrm{avar}}
	\global\long\def\MSE{\mathrm{MSE}}
	\global\long\def\OLS{\mathrm{OLS}}

	\global\long\def\sigerr{\sigma_{e}^{2}}
	\global\long\def\hatsigerr{\widehat{\sigma}_{e}^{2}}
	\global\long\def\bolSigmaf{\bolSigma_{\mbf}}
	\global\long\def\tilssigma{\widetilde{\sigma}^{2}}
	\global\long\def\hatssigma{\widehat{\sigma}^{2}}
	\global\long\def\ssigma{\sigma^{2}}
	\global\long\def\PLS{\mathrm{PLS}}
	\global\long\def\hatlambda{\widehat{\lambda}}
	\global\long\def\hatpi{\widehat{\pi}}
	
	\global\long\def\CRE{\mathcal{R}_{\bolSigma_{Y}}(\calB)}
	
	\global\long\def\CS{\calS_{Y\mid\mbX}}
	\global\long\def\hatsigma{\widehat{\sigma}}
	\global\long\def\hatdelta{\widehat{\delta}}
	\global\long\def\hatb{\widehat{b}}
	\global\long\def\tilb{\widetilde{b}}
	
	\newtheorem{lemma}{Lemma}
	\newtheorem{proposition}{Proposition}
	\newtheorem{theorem}{Theorem}
	\newtheorem{definition}{Definition}
	\newtheorem{example}{Example}
	
	\newcommand{\beqn}{\begin{equation*}}
	\newcommand{\eeqn}{\end{equation*}}
	
	\newcommand{\bea}{\begin{eqnarray}}
	\newcommand{\eea}{\end{eqnarray}}
	
	\newcommand{\bean}{\begin{eqnarray*}}
		\newcommand{\eean}{\end{eqnarray*}}
	
	\newcommand{\beq}{\begin{equation}}
	\newcommand{\eeq}{\end{equation}}
	
	\newcommand{\texthl}[1]{{ {\color{blue} #1}}}
	
	
	
	\section{Introduction} \label{sec:intro}
	
	Linear discriminant analysis (LDA) is one of the most popular classification method and a cornerstone for multivariate statistics \citep[e.g]{michie1994machine}. Classical LDA builds a linear classifier based on $p$-dimensional multivariate predictor $\mbX\in\mbbR^p$ to distinguish $K$ classes and to predict the class label $Y\in\{1,\ldots,K\}$. Despite its simplicity, LDA is shown to be very accurate on many benchmark datasets \citep{lim2000comparison,dettling2004bagboosting,hand2006classifier}. Moreover, LDA is easily interpretable and is thus often used as a visualizing tool for exploratory data analysis.

	In recent decades, the advancements in science and technology have enabled researchers to collect datasets with increasing sizes and complexity. Such datasets pose challenges to LDA. Four challenges that we tackle with this package are as follows. First, in research areas such as biology, genomics and psychology, we often have more predictors than samples. However, LDA is not applicable on these high-dimensional data, because sample covariance matrix becomes not invertible when the number of predictors exceeds the sample size. 
	
	Secondly, when we have a large number of predictors, variable selection is often desired such that we can obtain a sparse classifier involving only a small proportion of the variables. On one hand, \citet{fan2008high, bickel2008} showed in theory that variable selection is critical for accurate classification. On the other hand, sparse classifiers much easier to interpret in practice. However, LDA generally does not perform variable selection.

Thirdly, contemporary datasets often have complicated structure that renders the linear classifier in LDA inadequate. For example, in the presence of thousands of predictors, it may be inappropriate them to model all of them with the normal distribution. Moreover, research in neuroimaging, computational biology and personalized recommendation produces data in the form of matrices (2-way tensor) or tensors. The analysis of tensor datasets requires considerable modification to the vector-based LDA model. 

Last but not least, integrative analysis with multiple data sources are drawing researchers' attention recently. Co-existence of diverse data types, such as vector, matrix and tensor calls for more sophisticated models to integrate the information from them to improve classification accuracy. It is critical to model the dependence among different types of data to reduce the noise level in the data and improve prediction accuracy \citep{catch}.

Motivated by these challenges, many methods have been proposed to generalize LDA to datasets with high dimensions, non-normality and/or higher order predictors. In this package, we implement six methods that generalize LDA to contemporary complicated datasets. All of them are developed under models closely related to the LDA model, and penalties are imposed to achieve classification accuracy and variable selection in high dimensions. These methods include: 
\begin{enumerate}
\item Direct sparse discriminant analysis (DSDA): DSDA generalizes the classical LDA model to high dimensions when there are only two classes \citep{DSDA}. It formulates high-dimensional LDA into a penalized least squares problem.
\item Regularized optimal affine discriminant (ROAD): under the same model as DSDA, ROAD  fits a sparse classifier by minimizing the classification error under the $\ell_1$ constraint \citep{ROAD}.
\item Sparse optimal scoring (SOS) for binary problems: SOS is also developed under the LDA model \citep{Clemmensen}. It penalizes the optimal scoring problem \citep{hastie1994flexible}. We focus on its application in binary problems.
\item Semiparametric sparse discriminant analysis (SeSDA): SeSDA assumes a semiparametric model where data transformation can be applied to alleviate the non-normality. In practice, SeSDA finds the data-driven transformation and then performs model-fitting on the transformed data \citep{Mai2015ssda}. 
\item Multiclass sparse discriminant analysis (MSDA):  Instead of focusing on binary problems, MSDA considers the multiclass LDA model \citep{MSDA}. It takes note of the fact that the Bayes' rule can be estimated with minimizing a quadratic loss. To account for the multiclass structure, a group lasso penalty \citep{yuan2006model} is applied to achieve variable selection.

\item Covariate-adjusted tensor classification in high-dimensions (CATCH):
CATCH \citep{catch} is developed for tensor predictors. It takes advantage of the tensor structure to significantly reduce the number of parameters and hence alleviate computation complexity. 

\end{enumerate}

	\begin{table}[t]
		\centering
		\begin{tabular}{l|cccccc}
			\hline
			&DSDA&SOS&ROAD&SeSDA&MSDA&CATCH\\
			\hline
			Classes&Binary&Binary&Binary&Binary&Multi-class&Multi-class\\
			Data type&Vector&Vector&Vector&Vector&Vector&Tensor\\
			Model&LDA&LDA&LDA&SeLDA&LDA&TDA/CATCH\\
			Covariate adjustment&Yes&No&No&No&Yes&Yes\\
			\hline
		\end{tabular}
		\caption{Comparison of model settings between models. SOS was originally proposed to deal with both binary and multiclass problems, but we focus on binary problems in the package. Model SeLDA stands for Semi-parametric linear discriminant analysis, which is introduced in Section~\ref{Sec: SeLDA}. Model TDA/CATCH represents tensor discriminant analysis and covariate-adjusted tensor in high-dimensions, which are illustrated in Section~\ref{Sec: CATCH} and \ref{catchmodel}. }\label{tab:model}
	\end{table}

See Table~\ref{tab:model} for a comparison of these methods. Despite their different model assumptions and formulas, all of them have strong theoretical support and excellent empirical performance. We further note that they can be combined with covariate adjustment when multiple data sources are available. Our package $\tt TULIP$ integrates diverse discriminant analysis models and supportive functions to make it a convenient and well-equipped toolbox. It has several notable advantages. First, we not only include functions for model fitting, but also cross validation functions for easy control of the sparsity level, and prediction functions for the prediction of future observations. In addition, we provide covariate adjustment functions that efficiently remove heterogeneity in the predictors and combine information from covariates. Second, our package greatly facilitates the application of DSDA, ROAD and SeSDA for R users, as they do not have public R packages on CRAN outside ours. Third, although MSDA and SOS have been implemented in packages $\tt msda$ and $\tt sparseLDA$, we carefully modify their algorithms in our implementation to lower storage cost and/or speed up computation.


We acknowledge that many other efforts have been spent on topics closely related to that of our paper. On one hand, by now a large number of high-dimensional discriminant analysis methods have been developed. Some excellent examples include \citet{fan2008high, Tibshirani2002,Trendafilov2007, ROAD, Wu2009, cai2011constrained, Shao2011, Clemmensen, witten, Xu2015, Niu2015}. On the other hand, in the literature, many works study matrix/tensor regression and classification methods. Many of them impose low rank assumption \citep{Zhou2013, KoldaBader09Tensor, chi2012tensors, liu2017characterizing,CMDA, STDA, Zhong2015, Zeng2015}. All these methods have been reported to have great performance, but a comprehensive study of them is apparently out of the scope of our current paper.

The rest of this paper is organized as follows.	We start with a brief overview of discriminant analysis models in Section~\ref{Sec: models}. Model estimation and implementation details are discussed in Section~\ref{sec:para}. Section~\ref{sec: usage} contains instructions and examples on the usage of the package. A real data example is given in Section~\ref{sec: data} to confirm the numerical performance of methods in the package.

	\section{Discriminant Analysis Models and Bayes Rules}\label{Sec: models}
	
	\subsection{Bayes rule for classification}
	Recall that $Y\in\{1,\ldots,K\}$ is the categorical response (class indicator), and we use the generic $\mathcal{X}$ to denote the predictor and (potential) additional covariate. Specifically, $\mathcal{X}=\mbX\in\mbbR^p$ in classical multivariate discriminant analysis; $\mathcal{X}=\mbX\in\mbbR^{p_1\times\cdots\times p_M}$ in tensor discriminant analysis; and $\mathcal{X}=(\mbX,\mbU)$ in covariate-adjusted classification settings, where $\mbU\in\mbbR^q$ is additional covariates and $\mbX$ can be either vector or tensor. Our goal is to construct the optimal classifier to distinguish and predict $Y$ based on $\mathcal{X}$ under various settings. Denote $\pi_k=\Pr(Y=k)$ and $f_k$ as the conditional distribution of $\mathcal{X}$ within Class $k$ (e.g.~$f_k(\mathcal{X})=f(\mbX,\mbU\mid Y=k)$ is the joint distribution of $\mbX$ and $\mbU$ given $Y=k$, in presence of $\mbU$). The optimal classifier, often referred to as the Bayes rule, is thus
	\beq
	\delta(\bX)=\arg\max_k \{\log{\pi_k}+\log{f_k(\mathcal{X})}\}.
	\eeq
	The Bayes rule achieves the lowest classification error possible \citep{FHT01}. Therefore, it is our ultimate goal to estimate the Bayes rule. However, additional model assumptions are often needed for $f_k$ to ensure statistical and computational efficiency. 
	Consider the classical LDA setting of $\mbX\in\mbbR^p$ and $Y\in\{1,\dots,K\}$. To gain intuition, we often assume that within each class, the predictor follows a normal distribution with different means and a common covariance matrix. Then the Bayes rule is a linear function of $\mbX$ and can be straightforwardly estimated.

In the rest of this section, we discuss various statistical models that have been widely studied in the literature, along with the Bayes rules under these assumptions. Specifically, we review the classical LDA model, the semiparametric LDA model, and the tensor discriminant analysis model. We also discuss a general  framework for covariate adjustment. 
	 
	\subsection{The Linear Discriminant Analysis Model (LDA)}\label{Sec: LDA}

	Given a multivariate predictor $\mbX\in\mathbb{R}^p$ and $Y\in\{1,\ldots,K\}$, the LDA model assumes that $\mbX$ is normally distributed within each class, i.e,
	\begin{equation}\label{LDA}
	\mbX\mid(Y=k)\sim N(\bolmu_k,\bolSigma),\quad \Prob(Y=k)=\pi_k,\quad k=1,\ldots,K,
	\end{equation}
	where $\bolmu_k\in\mathbb{R}^p$ is the mean of $\mbX$ within class $k$, and $\bolSigma\in\mathbb{R}^{p\times p}$ is the common within class covariance matrix.

	Define $\bolbeta_k=\bolSigma^{-1}(\bolmu_k-\bolmu_1)$ for $k=1,\cdots,K$. The Bayes' rule turns out to be a linear function:  
	\begin{equation}\label{ldabayes}
	\widehat{Y}=\arg\max_{k}\Prob(Y=k\mid\mbX)=\arg\max_{k=1,\dots,K}\{\log\pi_{k}+\bolbeta_k^T(\mbX-\bolmu_k/2)\}.
	\end{equation}
	
The LDA model is simple yet elegant. All the parameters in this model have natural interpretations, while the Bayes rule has a nice linear form. An interesting fact about the Bayes rule in \eqref{ldabayes} is that it does not explicitly involve the $p^2$-dimensional parameter $\bolSigma^{-1}$. Instead, $\bolSigma^{-1}$ is only implicitly included in the discriminant directions $\bolbeta_k$. Moreover, it can be shown that the Bayes rule is equivalent to first reducing data to $\mbX^T\bolbeta_2,\ldots\mbX^T\bolbeta_K$ and then fitting the LDA model on the $(K-1)$-dimensional space. Therefore, to estimate the Bayes rule in high dimensions, our interest centers on the estimation of $\bolbeta_k$. We assume that $\bolbeta_k$'s are sparse with many elements being zero. Enforcement of this sparsity assumption will facilitate our estimation and naturally lead to variable selection.

Although the Bayes rule is derived under the somewhat restrictive normality and equal covariance assumptions, the discriminant directions $\bolbeta_k$ are still meaningful when data are non-normal, thanks to their geometric properties. It can be shown that, if we project $\mbX$ to $\bolbeta_k,k=1,\ldots,K$, the separation between classes is maximized over all possible sets of $K-1$ linear projections. Consequently, the LDA model is reasonably resistant to model misspecification. However, in some of the cases where the LDA model assumptions are severely violated, one can resort to more flexible models. For example, the quadratic discriminant analysis model \citep{jiang2015quda, fan2015quadro, LiandShao2015, Sun2015} relaxes the equal covariance assumption, while severely non-normal data can be modeled by the semiparametric model to be discussed in Section~\ref{Sec: SeLDA}.

	\subsection{Covariates Adjustment}\label{Sec:CA}

In many real-life problems, we have additional covariates along with the predictors. The covariates play two roles in the classification: it has predictive power on it own, and it also accounts part of the variation in the predictors. For example, in genomics studies, we record not only gene expression levels but also age and clinical measurements. In this case, we may view the gene expression levels as the high-dimensional predictor, and the age and clinical measurements as the covariates. We consider an LDA-type model to incorporate the covariates. In addition to the response $Y$ and the predictor $\bX$, we denote the covariates as $\bU\in\mathbb{R}^q$. We assume that
\begin{eqnarray}\label{CATCH_eq1}
	\mbU\mid(Y=k) & \sim & N(\bolphi_{k},\bolPsi), \label{CALDA.eq1}\\
	\mbX\mid(\mbU=\mbu,Y=k) & \sim & N(\bolmu_{k}+\bolalpha \mbu,\bolSigma),\label{CALDA.eq2}
	\end{eqnarray}
	where $\bolphi_{k}\in\mbbR^{q}$ is the within-class mean, $\bolPsi\in\mathbb{R}^{q\times q}$, $\bolPsi>0$ is the common within class covariance matrix of covariates, and $\bolalpha\in\mbbR^{p\times q}$ is the dependence of $\mbX$ on $\mbU$. We refer to this model as the covariate-adjusted LDA (CA-LDA) model. The CA-LDA model is conceptually similar to the CATCH model \citep{catch} for tensor, which is to be introduced in Section~\ref{Sec: CATCH}, but the CA-LDA model focuses on vector predictor $\mbX$ rather than tensor predictor. 
	
\begin{figure}[t!]
		\begin{center}
			\begin{tikzpicture}
			[scale=1.8,auto=right]
			\node[vertex][label=below: Covariates] (v1) at (1,4)  {$\mathbf{U}$};
			\node[vertex][label=Predictors] (v2) at (4,6.2)  {$\mathbf{X}$};
			\node[vertex][label=below: Class label] (v3) at (7,4)  {$\mbY$};
			
			\draw[-{Stealth[scale=1.8]}] (v1) to node [above,sloped]{{\Large $\bolalpha$}}(v2);
			\draw[-{Stealth[scale=1.8]}] (v1) to node [below,sloped]{Analogous to regression}(v2);
			\draw[-{Stealth[scale=1.8]}] (v2) to node [above,sloped] {{\Large $\bolbeta_2,\cdots,\bolbeta_K$}} (v3);
			\draw[-{Stealth[scale=1.8]}] (v2) to node [below,sloped] {LDA model} (v3);
			\draw[-{Stealth[scale=1.8]}] (v1) to node {{\Large $\bolgamma_2,\cdots,\bolgamma_K$}} (v3);
			\draw[-{Stealth[scale=1.8]}] (v1) to node [above]{LDA model} (v3);
			\end{tikzpicture}
		\end{center}
			\caption{\label{fig:covariate} Graphical illustration of the direct and indirect effects. The direct effect of covariate $\mbU$ on $Y$ follows classical discriminant analysis model measured by $\{\bolgamma_2,\ldots,\bolgamma_K\}$. Meanwhile, $\mbU$ also affects class label through affecting $\mbX$. Therefore we have $\widehat{Y}=f(\mbX, \mbU)$.
			}
\end{figure}
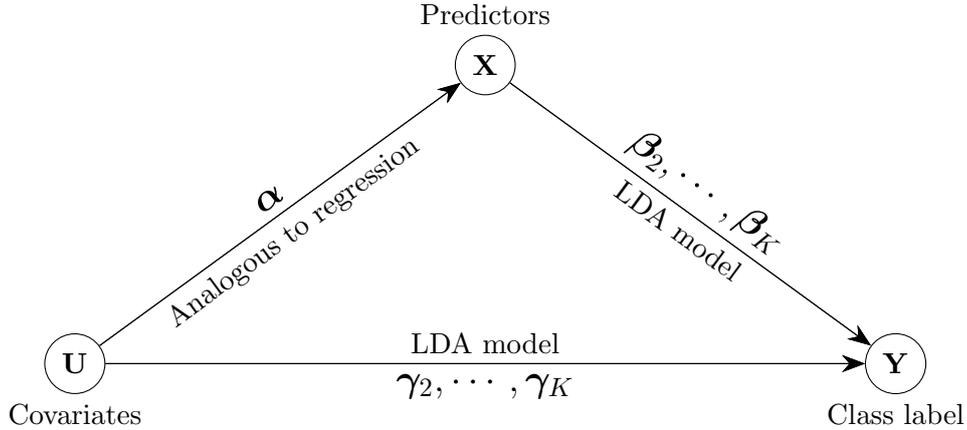

	Obviously, the CA-LDA model reduces to the LDA model in the absence of covariates. With the covariates, the CA-LDA model continues to have natural interpretations. Equation~\eqref{CALDA.eq1} indicates that $(\mbU,Y)$ marginally follow the LDA model. Equation~\eqref{CALDA.eq2} implies that the distribution of $\mbX$ not only depends on $Y$, but also $\mbU$ through mean dependence. Therefore, within each class, $\mbX$ is linked to $\mbU$ through a linear regression model, while, after we adjust for $\mbU$, $(\mbX,Y)$ follow the LDA model as well. See Figure~\ref{fig:covariate} for a graphical illustration of the relationship among $\mbX$, $\mbU$ and $Y$.

Under the CA-LDA model, the Bayes' rule is
	\begin{equation}
	\widehat{Y}=\arg\max_{k=1,\dots K}\left\{ a_k+\bolgamma_{k}^T\mbU+\bolbeta_k^T(\mbX-\bolalpha\mbU)\right\}
	\end{equation}
	where $\bolgamma_k=\bolPsi^{-1}(\bolphi_k-\bolphi_1)$, $\bolbeta_k=\bolSigma^{-1}(\bolmu_k-\bolmu_1)$ and $a_k=\log({\pi_k}/{\pi_1})-\frac{1}{2}\bolgamma_k^T(\bolphi_k+\bolphi_1)-\frac{1}{2}\bolbeta_k^T(\bolmu_k+\bolmu_1))$ is a scalar that does not involve $\mbX$ or $\mbU$. Throughout this paper, we assume that $\mbU$ is low-dimensional and does not need variable selection, but $\mbX$ is high-dimensional. In the presence of covariates, $\mbX$ needs to be first adjusted to $\mbX-\bolalpha\mbU$ before entering the Bayes rule. Similar to the LDA model, we assume that the coefficient of $\mbX-\bolalpha\mbU$, $\bolbeta_k$, is sparse.
	
	\subsection{The Semiparametric LDA model}\label{Sec: SeLDA}
	
Although LDA is reasonably resistant to model misspecification, we may still need more flexible models when data are heavily non-normal. The semiparametric linear discriminant analysis (SeLDA) model \citep{Lin2003} is proposed for this purpose. SeLDA assumes that there exists a set of strictly monotone univariate transformations $h_1,\ldots,h_p$ such that 
	\begin{equation}\label{sesda}
	(h_1(X_1),\cdots, h_p(X_p))\mid (Y=k)\sim N(\bolmu_k,\bolSigma).
	\end{equation} 
	
For identifiability, we further assume that all the diagonal elements in $\bolSigma$ are 1, and all elements in $\bolmu_1$ are 0. We also use the shorthand notation $h(\bX)=(h_1(X_1),\cdots, h_p(X_p))$. The transformation $h$ is assumed to be unknown and needs to be estimated from data. The SeLDA model assumes that the LDA model is true up to an unknown transformation. It has the same spirit as the well-known Box-Cox transformation, with which model assumptions are relaxed by proper data mapping.

It is easy to see that the LDA model is a special case of the SeLDA model, if we restrict $h(\mbX)=\mbX$. However, in the SeLDA model, we do not impose any parametric assumptions on $h$, which leads to great flexibility in practice. We further review a formula for $h_j$ that will facilitate its estimation. It can be shown that 
\beq\label{sesda.h}
h_j=\Phi^{-1}\circ F_{1j}=\Phi^{-1}\circ F_{kj}+\mu_{kj},
\eeq
 where $\Phi$ is the cumulative distribution function (CDF) of the standard normal random variable, and $F_{kj}$ is the CDF of $X_j$ within Class $k$. Equation~\eqref{sesda.h} will be used in Section~\ref{Sec: SeSDA}. The SeLDA model also amounts to assuming that the data follow the Gaussian copula model within each class \citep{wellner1997,HNW14,SemiCov}.

Although the SeLDA model requires much weaker conditions than the LDA model, it preserves many of the desirable properties. One of them is that the Bayes rule continues to be a linear function of the transformed data $h(\mbX)$:
	\begin{equation}\label{seldabayes}
	\widehat{Y}=\arg\max_{k=1,\dots,K}\{\log\pi_{k}+\bolbeta_k^T(h(\mbX)-\bolmu_k/2)\}.
	\end{equation}

Consequently, just as in the LDA model, when the dimension is high, we assume that $\bolbeta_k$ is sparse to allow accurate estimation.

	\subsection{Tensor Discriminant Analysis (TDA) and Covariate Adjustment} \label{Sec: CATCH}

\begin{figure}[H]
		\begin{minipage}{.5\textwidth}
			\centering
			\includegraphics[clip, trim=0.5cm 0.6cm 0.5cm 1cm, width=0.9\textwidth]{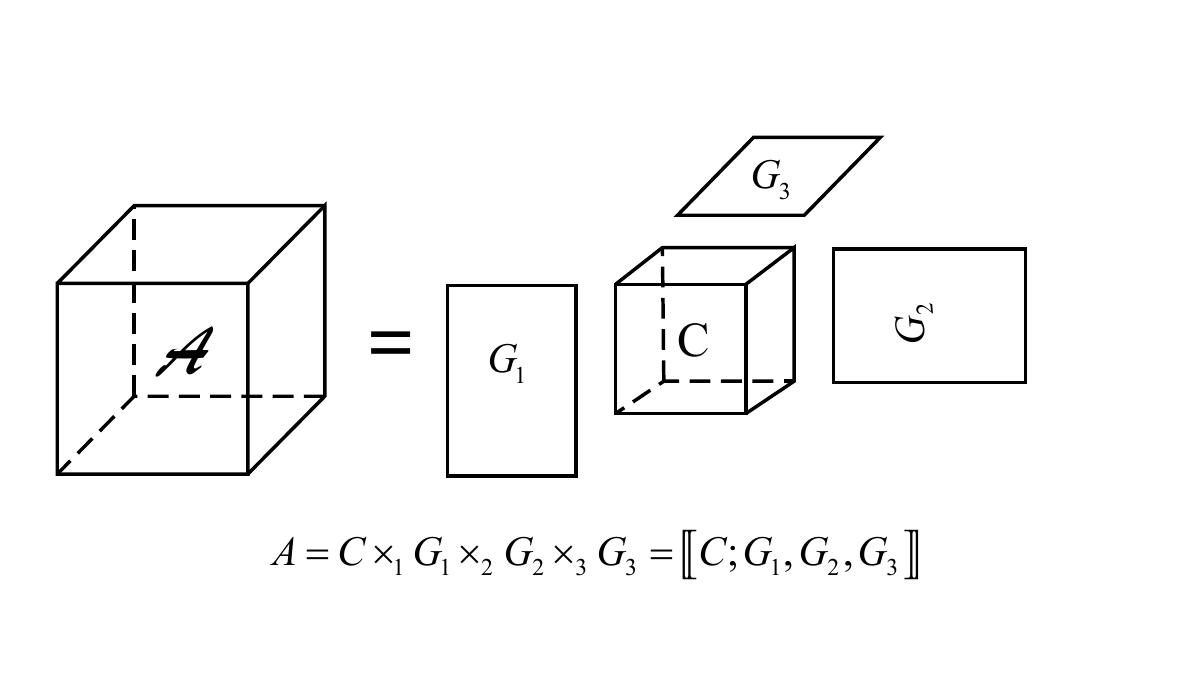}
			\caption{Tucker decomposition of tensor $\mbA$.}
			\label{fig:tucker}
		\end{minipage}
		\begin{minipage}{.5\textwidth}
		\centering
		\includegraphics[clip, trim=0cm 1.5cm 0cm 0.7cm, width=1.0\textwidth]{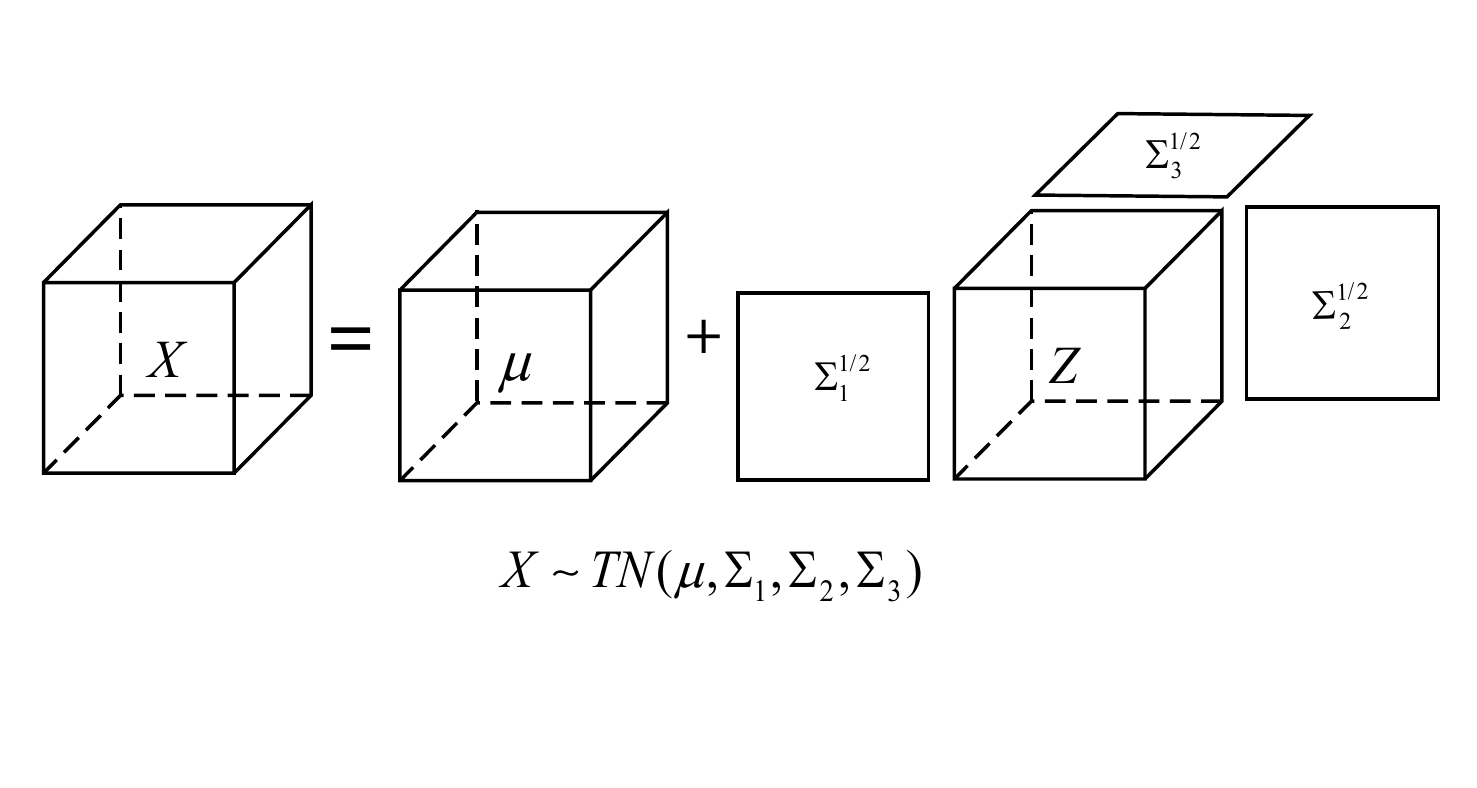}
		\caption{Tensor normal distribution}
		\label{fig:tn}
		\end{minipage}
	\end{figure}

	The tensor discriminant analysis (TDA) model is proposed for classification based on tensor predictors. We first briefly introduce some standard tensor notation \citep{KoldaBader09Tensor}. See Appendenx ~\ref{Append.A} for more rigorous definitions. An \emph{$M$-way tensor} is denoted by a multidimensional array $\mbA\in \mbbR^{p_{1}\times\cdots\times p_{M}}$ where $M\geq2$, $p_1,\ldots,p_M$ are all positive integers. We often need to multiply an $M$-way tensor $\mbC$ by $M$ matrices along each mode $\mbG_i,i=1,\ldots,M$, denoted by $\llbracket \mbC;\mbG_1,\ldots,\mbG_M\rrbracket$. For example, in Figure~\ref{fig:tucker} we obtain $\mbA=\llbracket \mbC;\mbG_1,\ldots,\mbG_3\rrbracket$ by multiplying a 3-way tensor $\mbC$ with matrices $\mbG_i$ along each mode. If $\mbG_i,i\ne m$ are identity matrices and $\mbG_m$ is a vector, then we write $\mbC\bar{\times}_{m}\mbG_m=\llbracket\mbC;\mbI,\ldots,\mbG_m,\ldots,\mbI\rrbracket$.

Further, we say a tensor $\mbX\in \mbbR^{p_{1}\times\cdots\times p_{M}}$ follows the tensor normal distribution $TN(\bolmu,\bolSigma_1,\ldots,\bolSigma_M)$ if it can be written as 
$$
\mbX=\bolmu+\llbracket\mbZ;\bolSigma_1^{1/2},\ldots,\bolSigma_M^{1/2}\rrbracket,
$$	
where $\mbZ\in \mbbR^{p_{1}\times\cdots\times p_{M}}$ has elements all independently standard normal, $\bolmu\in \mbbR^{p_{1}\times\cdots\times p_{M}}$ is the mean tensor, and $\bolSigma_m\in\mathbb{R}^{p_m\times p_m}$ are covariance matrices. See Figure~\ref{fig:tn} for an illustration.
	
Now we discuss the tensor discriminant analysis (TDA) model. Consider the $M$-way tensor predictor $\mbX\in\mathbb{R}^{p_1\times\cdots\times p_M}$ where $M\geq 2$ and class label $Y\in \{1,\ldots,K\}$. The TDA model assumes that

\begin{equation}\label{tda}
	\mbX\mid(Y=k)\sim TN(\bolmu_k,\bolSigma_1,\ldots,\bolSigma_M), \quad\Pr(Y=k)=\pi_k
	\end{equation}
	where $\bolmu_k\in\mathbb{R}^{p_1\times \cdots p_M}$, $\bolSigma_{m}\in\mathbb{R}^{p_m\times p_m}$ is the within-class mean, $\bolSigma_m>0$ is the common within-class covariance matrix along the $m$-th mode of the tensor, and $0<\pi_k<1$ is the prior probability for Class $k$. Compared to the LDA model, TDA utilizes the tensor normal distribution to model $\bX$ within each class. By taking advantage of the tensor structure, TDA drastically reduces the number of unknown parameters \citep{catch}. It can be seen that the TDA model requires $O(\sum_{m=1}^M p_m^2)$ parameters to model the dependence among $\mbX$. However, if we ignore the tensor structure and assume the LDA model on the vectorized version of $\mbX$, the covariance matrix has $O(\prod_{m=1}^Mp_m^2)$ parameters.

Under the TDA model, the Bayes' rule is
	\begin{equation}\label{bayestda}
	\widehat{Y}=\arg\max_{k=1,\dots K}\left\{ a_k+\langle\mbB_{k},\mbX\rangle\right\}
	\end{equation}
	where $\mbB_k=\llbracket \bolmu_k-\bolmu_1; \bolSigma_1^{-1},\ldots,\bolSigma_M^{-1}\rrbracket$,  and $a_k=\log({\pi_k}/{\pi_1})-\langle \mbB_k,\frac{1}{2}(\bolmu_k+\bolmu_1)\rangle$ is a scalar that does not involve $\mbX$. It can be seen that the Bayes rule is again a linear function in $\mbX$, with the linear coefficients $\mbB_k$. In high dimensions, we again impose the sparsity assumption by assuming that many elements in $\mbB_k$ are zeros.

	Similar to the vector case, when additional covariates are provided, the TDA model can be combined with covariate adjustment. \cite{catch} proposed the CATCH model for this purpose. In addition to $(Y,\mbX)$, we are given the covariates $\mbU\in\mathbb{R}^q$. The CATCH model assumes that
	\begin{eqnarray}
	\mbU\mid(Y=k) & \sim & N(\bolphi_{k},\bolPsi),\\
	\mbX\mid(\mbU=\mbu,Y=k) & \sim & TN(\bolmu_{k}+\bolalpha\bar{\times}_{(M+1)}\mbu,\bolSigma_{1},\dots,\bolSigma_{M}).
	\end{eqnarray}
where $\bolphi_k\in\mathbb{R}^q$ is the within-class mean of $\mbU$, $\bolPsi\in\mathbb{R}^{q\times q}$ is the within-class covariance of $\mbU$, and $\bolalpha\in\mathbb{R}^{p_1\times\cdots\times p_M\times q}$ characterizes the dependence of $\mbX$ on $\mbU$. The parameters in the CATCH model can be interpreted in the same way as the CA-LDA model in Section~\ref{Sec:CA}.

The Bayes' rule under the CATCH model is
	\begin{equation}\label{bayescatch}
	\widehat{Y}=\arg\max_{k=1,\dots K}\left\{ a_k+\bolgamma_{k}^T\mbU+\langle\mbB_{k},\mbX-\bolalpha\bar{\times}_{(M+1)}\bU\rangle\right\},
	\end{equation}
	where $\bolgamma_k=\bolPsi^{-1}(\bolphi_k-\bolphi_1)$, and $a_k=\log({\pi_k}/{\pi_1})-\frac{1}{2}\bolgamma_k^T(\bolphi_k+\bolphi_1)-\langle \mbB_k,\frac{1}{2}(\bolmu_k+\bolmu_1)\rangle$ is a scalar that does not involve $\mbX$ or $\mbU$. Similar to the TDA model, we assume that $\mbB_k$ is sparse in high dimensions, but impose no further sparsity assumptions on other parameters.

	\section{Methods}\label{sec:para}

	In this section, we formally introduce the six methods implemented by the package: DSDA, ROAD, SOS, SeSDA, MSDA and CATCH. 
	Throughout the rest of this paper, we denote $\widehat{\bolSigma}$ as the pooled sample covariance, $\widehat{\bolmu}_k$ as the within-class sample mean, $n$ as the sample size, and $n_k$ as the sample size in class $k$. All the methods involve a tuning parameter $\lambda>0$ that controls the amount of sparsity. Hence, when we refer to an estimate $\widehat{\bolbeta}$, it should be understood as $\widehat{\bolbeta}(\lambda)$, although we suppress $\lambda$ in most estimates for presentation convenience. We will discuss the tuning parameter in detail in Section~\ref{sec:tune}.	

	\subsection{Direct sparse discriminant analysis (DSDA)}

The direct sparse discriminant analysis (DSDA) is proposed for binary classification under the LDA model in \eqref{LDA}. Recall that our main interest is in estimating the coefficients $\bolbeta_k$ in the Bayes rule \eqref{ldabayes}. Because DSDA assumes that there are only two classes, it suffices to estimate $\bolbeta=\bolSigma^{-1}(\bolmu_2-\bolmu_1)$. In high dimensions, we assume that $\bolbeta$ is sparse. Let $y_i=-\frac{n_1}{n}$ if $Y_i=1$ and $y_i=\frac{n}{n_2}$ if $Y_i=2$. DSDA first solves the penalized least squares problem

	\begin{equation}\label{dsda}
	(\hatbolbeta^{\text{DSDA}},\widehat\beta_0^{\text{DSDA}})=\arg\min_{\bolbeta\in\mathbb{R}^p, \beta_0\in\mathbb{R}}\left\{ n^{-1}\sum_{i=1}^n(y_i-\beta_0-\mbX_i^T\bolbeta)^2+\lambda\sum_{j=1}^p\lvert\beta_j\rvert\right\},
	\end{equation}	
where $\lambda>0$ is the tuning parameter, $\sum_{j=1}^p\lvert\beta_j\rvert$ is the LASSO penalty \citep{Tibshirani1996}, and $\hatbolbeta^{\text{DSDA}}$ is our estimate for $\bolbeta$. Because of the LASSO penalty, $\hatbolbeta^{\text{DSDA}}$ is typically sparse. To estimate the Bayes rule, we further estimate the LDA model on the reduced data $\{Y_i,\mbX_i^T\hatbolbeta^{\text{DSDA}}\}_{i=1}^n$.

Numerical and theoretical studies show that DSDA consistently estimate the Bayes rule under mild conditions. Also, DSDA can be computed very efficiently, as \eqref{dsda} is a heavily-studied $\ell_1$ penalized least squares problem. Our implementation  utilizes $\tt glmnet$ to solve \eqref{dsda}.

\subsection{Regularized optimal affine discriminant (ROAD)}

Regularized optimal affine discriminnat (ROAD, \cite{ROAD}) is another binary penalized discriminant analysis method for high-dimensional data. ROAD estimates $\bolbeta$ by
	\begin{align}\label{ROAD}
	&\widehat{\bolbeta}^{\text{ROAD}}=\arg\min\bolbeta^T\widehat{\bolSigma}\bolbeta \\
	&\Vert \bolbeta\Vert_1\leq c, \bolbeta^T(\widehat{\bolmu}_2-\widehat{\bolmu}_1)/2=1.
	\end{align}
	We remark that \cite{wu2009sparse} independently proposed the $\ell_1$-Fisher's discriminant analysis method that closely resembles ROAD, but the developments of ROAD and the $\ell_1$-Fisher's discriminant analysis have different emphasis. ROAD clarifies several theoretical aspects of high-dimensional classification, while $\ell_1$-Fisher's discriminant analysis is developed for simultaneous testing for gene pathways. For simplicity, we focus on ROAD in what follows.  
	
In its optimization, the constraint of $\ell_1$-norm can be recast as a $\ell_1$-penalty with parameter $\lambda$. ROAD rewrites \eqref{ROAD} as
\beq\label{ROAD.opt}
\widehat{\bolbeta}^{\text{ROAD}}=\arg\min_{\bolbeta^T(\widehat{\bolmu}_2-\widehat{\bolmu}_1)/2=1}\bolbeta^T\widehat{\bolSigma}\bolbeta+\lambda\Vert\bolbeta\Vert_1
\eeq

The authors of ROAD proposed to solve \eqref{ROAD.opt} by replacing the nonconvex constraint with a quadratic penalty. However, we adopt a different approach to solve \eqref{ROAD.opt}. It is showed in \citet{Mai2013note} that the solution paths of DSDA and ROAD are equivalent. In other words, for any $\lambda>0$, there exists $\tilde\lambda>0$ such that $\widehat{\bolbeta}^{\text{DSDA}}(\lambda)\propto \widehat{\bolbeta}^{\text{ROAD}}(\tilde{\lambda})$. Because DSDA produces a solution path much faster than the original proposal of ROAD, we solve ROAD by first finding the solution path of DSDA for a range of $\lambda$, and then find each corresponding $\tilde{\lambda}$ to recover the solution path of ROAD.

	\subsection{Sparse optimal scoring (SOS) in binary problems}

	We also implement the successful discriminant analysis method, sparse optimal scoring (SOS, \cite{Clemmensen}). We focus on binary problems, where we are able to greatly improve the computation speed. For multiclass problems, SOS can be solved by the  $\tt R$ package \pkg{sparseLDA}.
	
In binary problems,	SOS creates a dummy variable $\mbY^{dm}\in\mbbR^{n\times 2}$ as a surrogate for the categorical response $Y$, where $Y_{ik}^{dm}=1\{Y_{i}=k\}$. Then SOS estimates coefficient by solving
	\begin{eqnarray}\label{SOS}
	\widehat\bolbeta^{\text{SOS}}&=&\arg\min_{\boltheta\in\mathbb{R}^2,\bolbeta\in\mathbb{R}^p}\{\Vert\mbY^{dm}\boltheta-\widetilde{\mbX}\bolbeta\Vert^2+\lambda\Vert\bolbeta\Vert_1\}, \nonumber\\
	&&\quad\mbox{ s.t $\frac{1}{n}\boltheta^T\mbY^{dm^T}\mbY^{dm}\boltheta=1, \boltheta^T\mbY^{dm^T}\mbY^{dm}1=0$},
	\end{eqnarray}
	where $\widetilde{\mbX}$ is the centered $\mbX$, and $\boltheta\in\mathbb{R}^{2}$ is the score for the two classes. SOS is a popular penalized discriminant analysis method because of its impressive empirical performance. It can be solved by iteratively minimizing the objective function in \eqref{SOS} over $\boltheta$ and $\bolbeta$.

	However, we take another approach to solve SOS with lower computation cost. \cite{Mai2013note} showed that $\widehat{\bolbeta}^{\text{SOS}}$ is closely related to the DSDA estimator defined in \eqref{dsda}. Let $\hat{\pi}_y=\frac{n_y}{n}$. We have that
\beq
\widehat{\bolbeta}^{\text{SOS}}(\lambda)=\sqrt{\hat{\pi}_1\hat{\pi}_2}\widehat{\bolbeta}^{\text{DSDA}}(\dfrac{\lambda}{\sqrt{\hat{\pi}_1\hat{\pi}_2}}).
\eeq
	
Therefore, to solve for $\widehat{\bolbeta}^{\text{SOS}}(\lambda)$, we first find $\widehat{\bolbeta}^{\text{DSDA}}(\dfrac{\lambda}{\sqrt{\hat{\pi}_1\hat{\pi}_2}})$ with DSDA, and rescale it to obtain the SOS solution. This approach avoids iteration between $\boltheta$ and $\bolbeta$, and is often faster than the original algorithm for SOS.

	\subsection{Semiparametric sparse discriminant analysis (SeSDA)}\label{Sec: SeSDA}

SeSDA \citep{Mai2015ssda} fits the SeLDA model in \eqref{sesda} for binary problems. It is expected to have better performance than DSDA when data are heavily non-normal. SeSDA has two steps. First, we find an estimate $\widehat{h}$ for the unknown function $h$. Second, we apply DSDA on the pseudo data $(\widehat{h}(\mbX),Y)$. In what follows, we focus on the estimation of $h$.

Two estimators have been proposed for $h$ based on \eqref{sesda.h}, the naive estimator and the pooled estimator. Without loss of generality, we assume that Class 1 has more observations than Class 2. Denote $\tilde{F}_{1j}$ as the empirical CDF of $X_j$ within Class 1.
To avoid infinity values at tails, we further Winsorize $\tilde{F}_{1j}$ to $\hat F_{1j}$, where
	
	 \[
	\hat{F}_{1j}(x)=\left\{
	\begin{array}{ll}
	1-1/n_1^2 &\textrm{ if } \tilde{F}_{1j}(x)>1-1/n_1^2\\
	\tilde{F}_{1j}(x) &\textrm{ if } 1/n_1^2\leq \tilde{F}_{1j}(x)\leq 1-1/n_1^2 \\
	1/n_1^2 &\textrm{ if } \tilde{F}_{1j}(x)<1/n_1^2.
	\end{array}
	\right.
	\]
	
	The naive estimator is shown to consistently estimate $h$, but in practice it is vulnerable to loss of efficiency, as it only utilizes one class of data. Therefore, the pooled estimator is proposed as a more efficient estimator. 
	
Similar to $\hat{F}_{1j}$, we denote $\hat{F}_{2j}$ as the empirical CDF of $X_j$ within Class 2 Winsorized at $(1/n_2^2,1-1/n_2^2)$. We first find an estimate for $\bolmu_{2j} $ as $\widehat{\bolmu}^{\text{(pool)}}_{2j}=\hat{\pi}_1\hat\mu_{2j}^{(1)}+\hat{\pi}_2\hat\mu_{2j}^{(2)}$, where $\hat\mu_{2j}^{(1)}=\frac{1}{n_2}\sum_{Y_i=2}\Phi^{-1}\circ \hat F_{1j}(X_{ij}), \hat\mu_{2j}^{(2)}=-\frac{1}{n_1}\sum_{Y_i=1}\Phi^{-1}\circ \hat F_{2j}(X_{ij})$. Then the pooled estimator for $h_j$ is
\beq
\widehat{h}_j^{\text{(pool)}}=\hat{\pi}_1\hat h_j^{(1)}+\hat{\pi}_2\hat{h}_j^{(2)},
\eeq	
where $\hat{h}_j^{(1)}=\Phi^{-1}\circ \widehat{F}_{1j}$ and $\hat{h}_j^{(2)}=\Phi^{-1}\circ \widehat{F}_{2j}+\hat{\mu}^{\text{(pool)}}_{2j}$.	The pooled estimator is usually more accurate than the naive estimator because it utilizes both classes to form an estimate for $h_j$.

%

	\subsection{Multiclass sparse discriminant analysis (MSDA)}\label{msdamodel}
	
Up to now, we have focused on binary classifiers. In this section, we discuss a multiclass classifier under the LDA model \eqref{LDA}. Assume that $K\ge 2$. By the Bayes rule \eqref{ldabayes}, we need to estimate the coefficients $\bolbeta_k=\bolSigma^{-1}(\bolmu_k-\bolmu_1), k=2,\ldots,K$. There is no need to estimate $\bolbeta_1$, as it is zero by definition. As in the binary problems, we continue to assume that the classifier is sparse in high dimensions, in the sense that only a few predictors are relevant to classification. However, this sparsity assumption has slightly different implication in multiclass problems. Note that, for any $X_j$, if any one of $\beta_{2j},\ldots,\beta_{Kj}$ is nonzero, $X_j$ is important for classification, as it helps with distinguishing between at least one pair of classes. Therefore, in order for an $X_j$ to be unimportant, we have to have $\beta_{2j}=\ldots=\beta_{Kj}=0$. In other words, the coefficients $\bolbeta_2,\ldots,\bolbeta_K$ has a group sparsity structure.

The multi-class sparse discriminant analysis (MSDA) has been proposed for fitting a sparse classifier under the context of interest. It takes note of the fact that, on the population level, we have

\begin{equation}
	(\bolbeta_2,\cdots,\bolbeta_K)=\arg\min_{\bolbeta_2,\cdots,\bolbeta_K}\sum_{k=2}^K\{\frac{1}{2}\bolbeta_k^T\bolSigma\bolbeta_k-(\bolmu_k-\bolmu_1)^T\bolbeta_k\}.
	\end{equation}

Therefore, in high dimensions, MSDA replaces the parameters with the sample estimates and impose the group sparsity structure through group lasso \citep{yuan2006model}. More specifically, MSDA estimates $\bolbeta$ by
	\begin{equation}\label{MSDA}
	(\widehat\bolbeta_2,\cdots,\widehat\bolbeta_K)=\arg\min_{\bolbeta_2,\cdots,\bolbeta_K}\sum_{k=2}^K\{\frac{1}{2}\bolbeta_k^T\hatbolSigma\bolbeta_k-(\hatbolmu_k-\hatbolmu_1)^T\bolbeta_k\}+\lambda\sum_{j=1}^P\parallel \bolbeta_{\cdot j}\parallel.
	\end{equation}
 The problem in \eqref{MSDA} can be solved by a blockwise coordinate descent algorithm \citep{MSDA} summarized in Algorithm \ref{alg:MSDA}. We refer to Algorithm~\ref{alg:MSDA} as the original MSDA algorithm. The $\tt R$ package $\tt msda$ implements such an algorithm. However, the original MSDA algorithm can be demanding on storage for high-dimensional data, because it requires the input of $\widehat{\bolSigma}\in\mathbb{R}^{p\times p}$. When $p$ is very large, the original MSDA algorithm can be practically inapplicable. Moreover, because of the sparse nature of $\bolbeta$, many elements in $\widehat{\bolSigma}$ are never used, and the calculation of them leads to unnecessary computation burden.
 
\begin{algorithm}[t]
		\begin{enumerate}
			\item Compute $\hatbolSigma$ and $\hatboldelta^k=(\hatbolmu_k-\hatbolmu_1)$, $k=1,2,\cdots, K$.
			\item Initialize $\hatbolbeta_k^{(0)}$ and compute $\tilbolbeta_k^{(0)}$ by $\tilbolbeta_{k,j}=\frac{\hat\delta_j^k-\sum_{l\neq j}\hat{\sigma}_{lj}\hat{\beta}_{kl}}{\hat{\sigma}_{jj}}$.	
			\item For steps $w=1,2,\ldots$, do the following until convergence:
			
			for each element $j=1,\ldots,p$,
			\begin{enumerate}
				\item Compute 
				\begin{equation}
				\hatbolbeta_{\cdot j}^{(w)}=\tilbolbeta_{\cdot j}^{(w-1)}(1-\frac{\lambda}{\parallel \tilbolbeta_{\cdot j}^{(w-1)}\parallel})_{+};
				\end{equation}
				\item Update
				\begin{equation}\label{algmsdaupdate}
				\tilde{\beta}_{kj}=\frac{\hat{\delta}_j^k - \sum_{l\neq j}\hat{\sigma}_{lj}\hat{\beta}_{kl}^{(w)}}{\hat{\sigma}_{jj}}.
				\end{equation}
			\end{enumerate}
			\item At convergence, output $\bolbeta_{k}$.
		\end{enumerate}
		\caption{\label{alg:MSDA} Algorithm for MSDA}
	\end{algorithm}

 Therefore, in our implementation we modify the original MSDA algorithm for lower storage and computation cost for high-dimensional data. Note that $\widehat{\bolSigma}$ is only used in updating rule \eqref{algmsdaupdate}. We take advantage of two properties of this  updating rule \eqref{algmsdaupdate}. First, given the natural element-wise property of coordinate descent algorithm, only the $j$-th column of covariance matrix $\hatbolSigma_{\cdot j}$ is needed in each iteration. The full covariance matrix is never used during the computation process. Therefore, it is not necessary to store the huge covariance matrix. Secondly, a large number elements of $\hatbolbeta$ are exactly 0. Hence among the column $\hatbolSigma_{\cdot j}$, we only need to compute the rows corresponding to the nonzero coefficients.
	These facts motivate us to develop the modified MSDA algorithm. The modified MSDA algorithm is largely identical to the original algorithm, but with two important distinctions. On one hand, in Step 1 we only require the input of $\widehat{\boldelta}^k$ but not $\widehat{\bolSigma}$. On the other hand, Step 3(b) in \eqref{algmsdaupdate} is replaced with
	\begin{equation}\label{algMmsdaupdate}
	\tilde{\beta}_{kj}=\frac{(n-K)\hat\beta_j^k-\sum_{l\neq j}\hat\beta_{kl}^{(m)}(\sum_{k=1}^K[\sum_{i\in\mbT_k}(X_{il}-\mu_{kl})(X_{ij}-\mu_{kj})])}{\sum_{k=1}^K[\sum_{i\in\mbT_k}(X_{ij}-\mu_{kj})^2]},
	\end{equation}
	where $\mbT_k=\{i:y_i=k\}$. By doing so, we avoid the storage and the computation of the full matrix of $\hatbolSigma$. In computing \eqref{algMmsdaupdate}, we further use three tricks to speed up the computation. Firstly, we calculate and store all diagonal elements in the covariance matrix as they will be called multiple times.
	Secondly, we keep the indexes of nonzero elements in $\mbT_k$ and update it every time we observe a new nonzero element. Hence we do not need to check all elements to locate the nonzero ones in each iteration. Thirdly, we update equation \eqref{algMmsdaupdate} by only computing elements corresponding to the nonzero indexes in $\mbT_k$. With these three tricks, the modified algorithm reduces the space complexity from $O(p^2)$ to $O(p)$, and is also faster than the original algorithm for large $p$.

	\subsection{Covariate-adjusted tensor classification in high dimensions (CATCH)}\label{catchmodel}

When $\mbX$ is a tensor instead of a vector, we need to fit the TDA model or the CATCH model (in presence of covariates) for better efficiency and accuracy. \cite{catch} proposed the CATCH method to fit both models, but in this section we focus on the CATCH method on the TDA model, where there is no covariate. The inclusion of covariates will be discussed in Section~\ref{sec:covlam}.

Recall that, under the TDA model, we aim to estimate the parameters $\mbB_k=\llbracket \bolmu_k-\bolmu_1;\bolSigma_1^{-1},\ldots,\bolSigma_M^{-1}\rrbracket$. We first rewrite $\mbB_k$ as solutions to estimating equations:
$$
(\mbB_2,\ldots,\mbB_K)=\arg\min_{\mbB_2,\ldots,\mbB_K}\sum_{k=2}^K\left(\langle\mbB_k,\llbracket\mbB_k;\bolSigma_{1},\dots,\bolSigma_{M}\rrbracket\rangle-2\langle\mbB_k,\bolmu_{k}-\bolmu_{1}\rangle\right),
$$
where for two $M$-way tensors $\mbA,\mbC$, $\langle\mbA,\mbC\rangle=\sum_{j_1\cdots j_M}a_{j_1\cdots j_M}c_{j_1\cdots j_M}$ is the inner product of two tensors. To estimate $\mbB_k$, we find the within-class sample mean $\hatbolmu_k$ as the estimate for $\bolmu_k$, and moment-based unbiased estimators $\hatbolSigma_{m}$ for $\bolSigma_{m}$; see the formulas in Appendix~\ref{Appendix.C}. We further add the group LASSO penalty for variable selection. Therefore, CATCH solves the following problem:
	\begin{equation}\label{TLDA-formula}
	\min_{\mbB_2,\ldots,\mbB_K}\left[\sum_{k=2}^K\left(\langle\mbB_k,\llbracket\mbB_k;\hatbolSigma_{1},\dots,\hatbolSigma_{M}\rrbracket\rangle-2\langle\mbB_k,\hatbolmu_{k}-\hatbolmu_{1}\rangle\right)
	+\lambda\sum_{j_{1}\dots j_{M}}\sqrt{\sum_{k=2}^{K}b_{k,j_{1}\cdots j_{M}}^2}\right].
	\end{equation}
	
CATCH can be solved by a coordinate descent algorithm with an explicit updating formula in each iteration.

	%
	%

	\subsection{Covariates adjustment}\label{sec:covlam}
		
	When we have additional covariates $\mbU$, the CA-LDA model or the CATCH model should be fitted. Whether $\mbX$ is a vector or a tensor, a key step for the covariate adjustment is the estimation of $\bolalpha$, the dependence of $\mbX$ on $\mbU$. We use the maximum likelihood estimator (MLE). Denote $\overline{\mbU}_k$ as the sample mean of $\mbU$ within class k and $\overline{\mbX}_k$ as the sample mean of $\mbX$ within class k. Define group-wise centered data $\widetilde{\mbX}_i=\mbX_i-\overline{\mbX}_{Y_i}$, $\widetilde{\mbU}_i=\mbU_i-\overline{\mbU}_{Y_i}$.

		For vector-variate $\mbX_i\in\mbbR^p$, we adjust for covariate $\mbU$ by $\mbX_i-\hatbolalpha\mbU_i$, where $\hatbolalpha\in\mbbR^{q\times p}$ is the MLE,
	\beq
	\hatbolalpha=(\widetilde{\mbU}^T\widetilde{\mbU})^{-1}\widetilde{\mbU}^T\widetilde{\mbX}.
	\eeq  	

For tensor-variate $\mbX_i\in\mbbR^{p_1\times\cdots\times p_M}$, we let $\bolalpha_{j_1\cdots j_M}\in\mathbb{R}^{q}$ be the regression coefficient of univariate $X_{i,j_1\cdots j_M}$ on multivariate $\mbU_i\in\mbbR^q$. Then the MLE for $\bolalpha_{j_1\cdots j_M}$ is $\widehat{\bolalpha}_{j_1\cdots j_M}=(\widetilde{\mbU}^T\widetilde{\mbU})^{-1}\widetilde{\mbU}^T\widetilde{X}_{j_1\cdots j_M}$, which can be expressed more explicitly as,
	 	\beq\label{alpha.MLE1}
	 	\widehat{\bolalpha}_{j_1\cdots j_M}=\left\{ \sum_{k=1}^{K}\sum_{Y_i=k}(\mbU_{i}-\overline{\mbU}_{k})(\mbU_{i}-\overline{\mbU}_{k})^T\right\} ^{-1}\left\{ \sum_{k=1}^{K}\sum_{Y_i=k}(\mbU_{i}-\overline{\mbU}_{k})(X_{i,j_1\cdots j_M}-\overline{X}_{k,j_1\cdots j_M})\right\}.
	 	\eeq
	 	
Afterwards, the ensemble of all $\widehat{\bolalpha}_{j_1\cdots j_M}$, $\widehat{\bolalpha}$, is our estimator for $\bolalpha$. The covariate-adjusted predictor is then obtained as $\mbX_i-\widehat{\bolalpha}\bar{\times}_{M+1} \mbU_i$.

\subsection{Selection of the tuning parameter}\label{sec:tune}

We recommend selecting the tuning parameter in all methods by cross validation, which is implemented in our package as supportive functions for most of the methods. In cross validation, a sequence of potential tuning parameters is supplied. For each candidate tuning parameter $\lambda$, the dataset is random split into $L$ folds. 
Then we fit $L$ classifiers, each of which is fitted on $L-1$ folds of the data and validated on the remaining one fold. The average validation error rate of the $L$ classifiers is used as a measurement of the performance of the corresponding $\lambda$. The $\lambda$ with the smallest average validation error is used in our final model fitting.

If desired, our package can automatically generate a sequence of tuning parameters for all the methods.  They will first compute the smallest $\lambda$ that shrinks all coefficients to zero; this value is taken as the upper bound of tuning range. Then the upper bound is multiplied by a small number to generate the lower bound. Finally,  a sequence of tuning parameters is uniformly generated between the lower and the upper bound.

	\section{Using the R Package}\label{sec: usage}

	The R package \textbf{TULIP} provides user-friendly functions to fit discriminant analysis model and perform predictions on vector and tensor data. The package can be downloaded through \texthl{cran link} or install in \proglang{R} through \code{install.packages('TULIP')}. In installing package, the pre-required packages \pkg{MASS} for LDA model fitting, packages \pkg{Matrix} \citep{Matrix} and \pkg{tensr} \citep{tensr} for matrix and tensor operations, and the package \pkg{glmnet} for LASSO are also automatically installed. Users do not need to install them separately. To guarantee higher computation efficiency of the package, core algorithms of MSDA and CATCH are implemented in Fortran, which have already been compiled and can also be used directly.
	
	Among all the six methods, there is always a tuning parameter $\lambda$ to control the size of sparsity. On the implementation aspect, MSDA and CATCH also have parameter \code{dfmax} to limit the number of selected variables and will only return the solutions with number of non-zero elements less than \code{dfmax}. Furthermore, MSDA has a \code{model} option to specify version of implementation between multi.original and multi.modified. The methods are summarized in Table~\ref{tab:modelsum}.
	
	\begin{table}[!t]
		\centering
			\begin{tabular}{ll|ccccc}
				\hline
				&Parameters&$\lambda$&\code{dfmax}&\code{model} option&Covariate&Cross validation\\
				\hline
				DSDA&Binary vector &\checkmark&&&\checkmark&\checkmark\\
				ROAD&Binary vector &\checkmark&&&\\
				SOS&Binary vector &\checkmark&&&\\
				SeSDA&Binary vector &\checkmark&&&&\checkmark\\
				MSDA&Multi-class vector &\checkmark&\checkmark&\checkmark&\checkmark&\checkmark\\
				CATCH&Multi-class tensor &\checkmark&\checkmark&&\checkmark&\checkmark\\
				\hline
			\end{tabular}
		\caption{Method description and major parameters. Penalty parameter $\lambda$ controls the size of $\ell_1$-penalty. Parameter \code{dfmax} limits the maximum number of non-zero variables. Parameter \code{model} specifies the version of implementation for MSDA.}
		\label{tab:modelsum}
	\end{table}
	
	The functions in the package consists of two parts. One part contains core functions which generate solution paths of all the methods, including functions \fct{dsda}, \fct{road}, \fct{sos}, \fct{SeSDA}, \fct{msda} and \fct{catch}. Since binary classification can be regarded as a special case of multi-class problems, we also embedded DSDA into \fct{msda} function. See Section~\ref{sec:core} for details.
	The other part includes supportive functions to perform covariate adjustment, prediction, cross validation and handle some special cases.

	To illustrate how to use the functions, we first simulate a binary vector data set named \code{dat.vec} with dimension $p=500$ and sample size $n_k=75$. In the data set, we have $\mbX_i\mid (Y_i=k) \sim N(\bolmu_k,\bolSigma)$, where $\bolmu_1=0$, $\bolmu_2=\bolSigma\bolbeta$, $\sigma_{ij}=0.3$ if $i\neq j$ and $\sigma_{ii}=1$, $\bolbeta_j=0.5$ for $1\leq j \leq 10$ and $\bolbeta_j=0$ otherwise. We further generate a testing data set with sample size 1000 from the same distribution. Variables in \code{dat.vec} is summarized in Table~\ref{tab:datvec}.  Data set \code{dat.vec} can be simulated by code
	\begin{CodeChunk}
		\begin{CodeInput}
R> dat.vec<-sim.bi.vector(1000)
		\end{CodeInput}
	\end{CodeChunk}

%

		\begin{table}[H]
		\centering
		\begin{tabular}{ccc}
			\hline
			Variable&Type&Dimension\\
			\hline
			x&matrix&$150\times 500$\\
			y&vector&150\\
			testx&matrix&$1000\times 500$\\
			testy&vector&1000\\
			\hline
		\end{tabular}
		\caption{Data set dat.vec.}
		\label{tab:datvec}
	\end{table}

	Moreover, we include two real data sets, GDS1615 and colorimetric sensor array data set, in the package to demonstrate usage of the functions. Data set GDS1615 \citep{Burczynski2006} is a vector data set where observations belong to three classes. The original data set contains 127 observations and 22283 variables. Package \pkg{msda} preprocessed the data by computing F-test statistics of each variable \citep{MSDA}, whose definition is in appendix. Hence only 127 variables are kept in the data set. Colorimetric sesor array data (CSA) was used to show the performance of discriminant analysis method \citep{Zhong2015}. It records information of chemical dyes after exposed to volatile chemical toxicants to identify their classes. It contains 147 observations in 21 classes. For each observation, the predictor is a $36\times 3$ matrix. We include two conditions in our dataset, but focus on the Immediately Dangerous to Life or Health (IDLH) condition. 
	
	\subsection{Core functions}\label{sec:core}
	
	\noindent {\bf Function \fct{dsda}}
	
	The following code shows an example of utilizing DSDA. Given the data set \code{dat.vec}, we fit DSDA on $\{\mbX, Y\}$ by specifying the tuning range of parameter $\lambda$ to be a sequence between $[0.005, 0.3]$. Hence the function will generate a solution path. Next, we apply \fct{predict} function on the model and obtain the prediction for each $\lambda$ and error rate. In the example, we report the minimum error rate and corresponding parameter value.

\begin{CodeChunk}
\begin{CodeInput}
R> obj <- dsda(dat.vec$x, y=dat.vec$y, lambda=seq(0.005, 0.3, length.out=20))
R> pred <- predict(obj, dat.vec$testx)
R> err<- apply(pred, 2, function(x){mean(x!=dat.vec$testy)})
R> print(min(err))
\end{CodeInput}
\begin{CodeOutput}
[1] 0.111
\end{CodeOutput}
\begin{CodeInput}
R> print(obj$lambda[which.min(err)])
\end{CodeInput}
\begin{CodeOutput}
[1] 0.02052632
\end{CodeOutput}
\end{CodeChunk}

If one wishes, \fct{dsda} can also be used in a more automatic way. On one hand, it can be called without supplying a sequence of value for tuning parameter. The function will automatically generate a sequence based on data. On the other hand, the prediction can be performed along with model fitting if testing data is supplied. The function \fct{dsda} will produce the prediction error on the testing data corresponding to each tuning parameter. See the following example.

\begin{CodeChunk}
\begin{CodeInput}
R> obj <- dsda(dat.vec$x, y=dat.vec$y,testx=dat.vec$testx)
R> err <- apply(obj$pred, 2, function(x){mean(x!=dat.vec$testy)})
R> print(min(err))
\end{CodeInput}
\begin{CodeOutput}
[1] 0.107
\end{CodeOutput}
\begin{CodeInput}
R> print(obj$lambda[which.min(err)])
\end{CodeInput}
\begin{CodeOutput}
[1] 0.03180946
\end{CodeOutput}
\end{CodeChunk}

\begin{figure}[!t]
		\centering
		\includegraphics[width=0.7\textwidth]{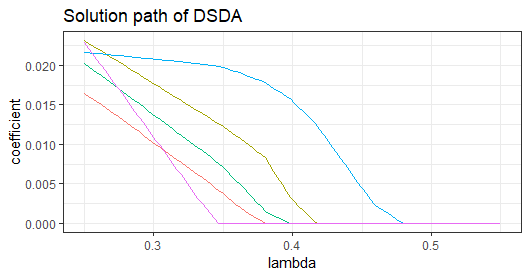}
		\caption{Solution path of five selected variables in a DSDA model.}
		\label{fig:dsdasp}
	\end{figure}

	Figure~\ref{fig:dsdasp} shows a solution path of DSDA model. As parameter $\lambda$ increases, more coefficients will be shrunken towards 0. In addition, DSDA can also integrate the covariate adjustment, model fitting and prediction. The usage is similar to the function \fct{catch}, and we do not give a separate example here to avoid redundancy.

		\noindent {\bf Function \fct{SeSDA}}
	
	Function \fct{SeSDA} fits a semiparametric sparse discriminant analysis model on the input vector data. The simulated data \code{dat.vec} follows normal distribution within each class. We take an exponential transformation on it to violate the normality assumption. The following example shows that \fct{SeSDA} achieves error rates 11\%. However, if we directly apply DSDA on the data set, the minimum error rate is as high as 15.8\%. Therefore, the preprocessing of SeSDA can indeed help to improve performance under this scenario.

\begin{CodeChunk}
\begin{CodeInput}
R> x <- exp(dat.vec$x)
R> testx <- exp(dat.vec$testx)
R> obj.SeSDA <- SeSDA(x, y=dat.vec$y)
R> pred.SeSDA <- predict(obj.SeSDA, testx)
R> err <- apply(pred.SeSDA, 2, function(x){mean(x!=dat.vec$testy)})
R> min(err)
\end{CodeInput}
\begin{CodeOutput}
[1] 0.11
\end{CodeOutput}
\end{CodeChunk}
	
\begin{figure}[!t]
\centering
\includegraphics[width=1.0\textwidth]{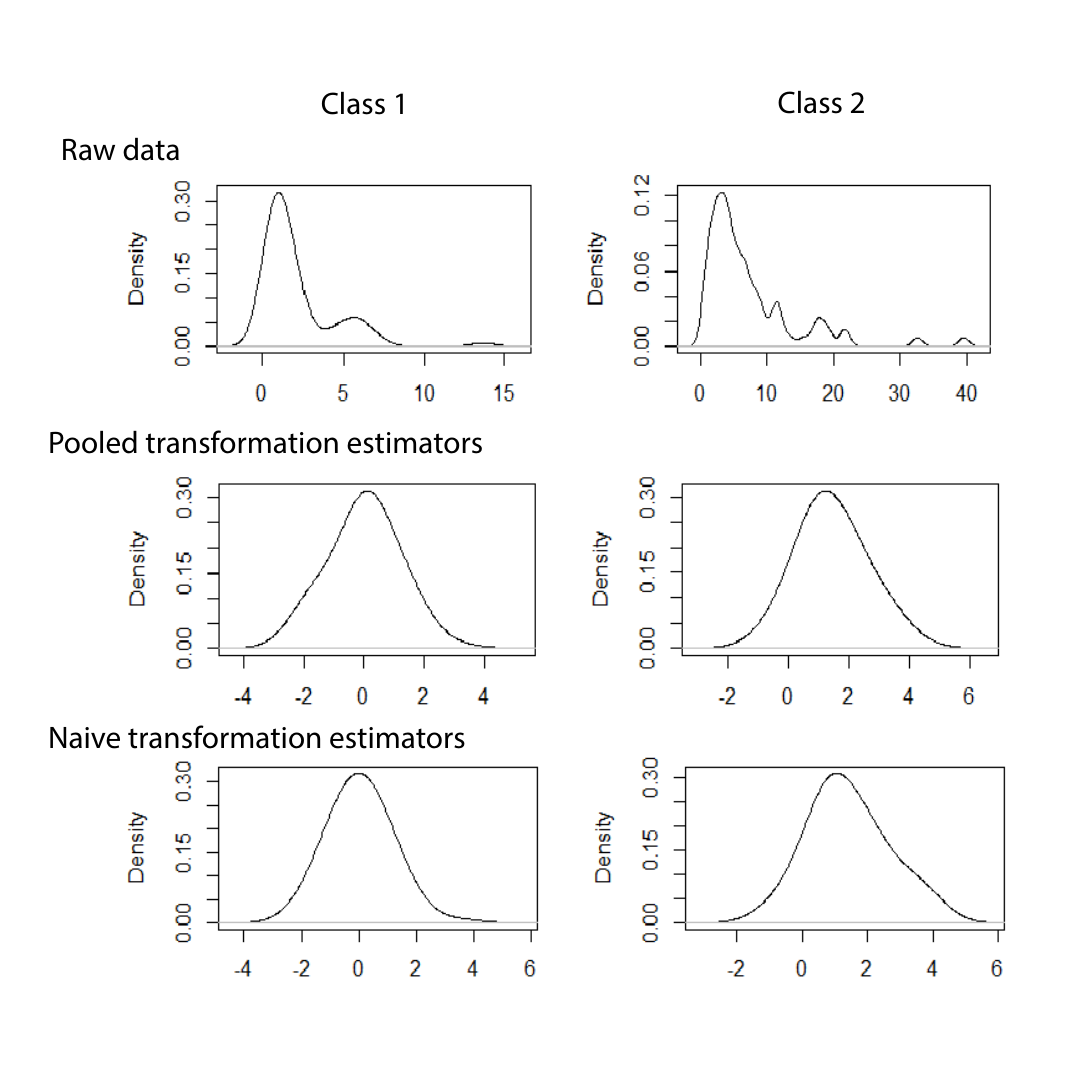}
\caption{The distribution of the 1st variable in simulated data set among two classes before transformation and after transformation. The top row is before transformation. The second row is after pooled transformation. The bottom row is after na\"{i}ve transformation.}
\label{fig:sesda}
\end{figure}

Further, Figure~\ref{fig:sesda} shows how the distribution of the first variable changes after transformation. It is clear that both pooled and na\"{i}ve transformatins result in approximately normal distribution.

	\noindent {\bf Functions \fct{ROAD} and \fct{SOS}}
	
\begin{figure}[!t]
		\centering
		\includegraphics[width=0.7\textwidth]{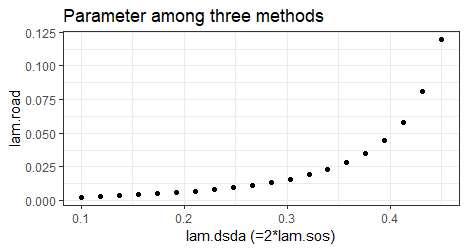}
		\caption{Parameters in ROAD vs. Parameters in DSDA. Notice that the parameters in DSDA are double those of SOS.}
		\label{fig:param}
	\end{figure}

	Functions \fct{ROAD} and \fct{SOS} can generate equivalent solution paths as ROAD \citep{ROAD} and SOS \citep{Clemmensen} methods on binary vector data, respectively. Both of the two models are fit by calling \fct{dsda} function. Compared to the original package for SOS, \pkg{sparseLDA}, our implementation is usually faster, especially when a solution path or parameter tuning is needed. For example, to fit a solution path with 10 possible values of $\lambda$s on a toy example with $p=40$, our implementation reduces the computation time by half compared to \pkg{sparseLDA}. An example of fitting ROAD and SOS model is as follows. The \code{lambda}s passed into \fct{ROAD} and \fct{SOS} functions will be directly used by \fct{dsda} function. The \code{lambda}s returned by the two functions are their corresponding parameters in ROAD and SOS model, respectively. Figure~\ref{fig:param} shows the relationship between the $\lambda$'s that generate the same solution.

\begin{CodeChunk}
\begin{CodeInput}
R> obj.dsda <- dsda(dat.vec$x, y=dat.vec$y, lambda=seq(0.1, 0.5, length.out=20))
R> obj.road <- ROAD(dat.vec$x, y=dat.vec$y, lambda=seq(0.1, 0.5, length.out=20))
R> obj.sos <- SOS(dat.vec$x, y=dat.vec$y, lambda=seq(0.1, 0.5, length.out=20))
\end{CodeInput}
\end{CodeChunk}

	\noindent {\bf Function \fct{msda}}

	The function \fct{msda} provides an interface to fit MSDA. Similarly to \fct{dsda}, without specification of possible values of $\lambda$, the function will automatically generate a sequence of $\lambda$s. Function \fct{msda} can also perform predictions when testing data is supplied and make adjustments on covariates when covariates exist. 	We apply \fct{msda} on GDS1615 data set to as a demonstration. We report the minimum training error, its corresponding parameter value and the number of non-zero variables selected by the model.

\begin{CodeChunk}
\begin{CodeInput}
R> data(GDS1615)
R> x <- GDS1615$x
R> y <- GDS1615$y
R> set.seed(123456)
R> teindex <- c(sample(which(y==1), sum(y==1)/3), sample (which(y==2), 
+  sum(y==2)/3), sample(which(y==3), sum(y==3)/3))
R> obj <- msda(x[-teindex, ], y=y[-teindex], testx=x[teindex, ])
R> err <- apply(obj$pred, 2, function(x){mean(x!=y[teindex])})
R> min(err)
R> paste(min(err), obj$lambda[which.min(err)], obj$df[which.min(err)] )
\end{CodeInput}
\begin{CodeOutput}
[1] "0.04878049 1.446872 19"
\end{CodeOutput}
\end{CodeChunk}

\begin{figure}[t]
	\centering
	\includegraphics[width=0.7\textwidth]{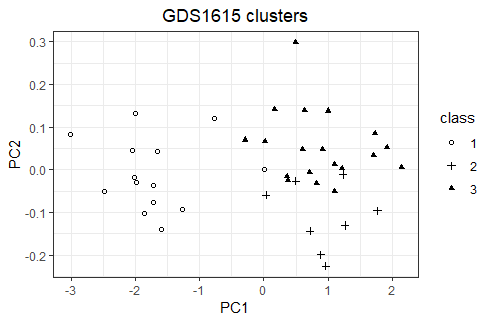}
	\caption{The GDS data projected onto the two principle components of $\mbX\bolbeta$.}
	\label{fig:msdagds}
\end{figure}

If one wishes to visualize the discriminant effect, plots of projections on the discriminant coefficients is helpful. A principle component analysis is also optional to show the classification even more clearly.
For illustration, we perform principle component analysis on $\mbX\bolbeta$ where $\bolbeta=\{\bolbeta_2, \bolbeta_3\}$ is the discriminant coefficient. The scatter plot on the two principle components is shown in Figure~\ref{fig:msdagds}. It is clear to see that the three classes are separated well.

We also note that \fct{msda} has an argument \code{model} that can be specified by users to use different algorithms in MSDA. The options for \code{model} include  \code{binary}, \code{multi.original} and \code{multi.modified}. The option \code{binary} can only be used in binary problems. If selected, MSDA is solved by DSDA, which gives the same solution with usually less time. However, using this option in multi-class problems will result in an error. The option \code{multi.original} indicates that MSDA is solved by the original algorithm, which requires the calculation of the full covariance matrix. When the dimension is low, \code{multi.original} is often efficient. The option \code{multi.modified}, on the other hand, solves MSDA with the modified algorithm, where only part of the covariance matrix is calculated in each iteration. This option allows MSDA to be applicable in much higher dimensions. Also, when \code{multi.modified} is selected, we suggest using relatively larger tuning parameters to account for the high dimensionality. If unspecified, \code{model} is set to be \code{binary} in binary problems. If the response variable is multi-class, the function will call multi.original implementation for $p\leq2000$ and multi.modified implementation for $p>2000$.

%
%

	\noindent {\bf Function \fct{catch}}
	
	To illustrate usage of function \code{catch}, we first simulate a data set named \code{dat.ten} with tensor predictors $\mbX_i\in\mbbR^{10\times 10\times 10}$ and covariates $\mbU_i\in\mbbR^2$. The data is simulated from model $\mbX_i\mid (Y_i=k) \sim TN(\bolmu_k,\bolSigma_1,\bolSigma_2,\bolSigma_3)$ where $\bolmu_1=0$, $\bolmu_2=\llbracket \bolbeta;\bolSigma_1,\bolSigma_2,\bolSigma_3\rrbracket$, $\bolSigma_j=\mbI$ for $j=1,2,3$, $\bolbeta_{[1:2,1:2,1:2]}=0.8$ and 0 otherwise. 
	Let $\mbU_i\mid (Y_i=k)\sim N(\bolphi_k,\bolpsi)$ where $\bolphi_1=0$, $\bolphi_2=(0.3, 0.3)$ and $\bolpsi=\mbI$. The connection between $\mbX$ and $\mbU$ is measured by $\bolalpha\in\mbbR^{10\times 10\times 10\times 2}$ and $\bolalpha_{[1:5,1:5,1:5,1]}=1$ and 0 otherwise. Variables in \code{dat.ten} are summarized in Table~\ref{tab:datten}.

 Data set \code{dat.vec} can be simulated by code
\begin{CodeChunk}
	\begin{CodeInput}
R> dat.ten<-sim.tensor.cov(1000)
	\end{CodeInput}
\end{CodeChunk}


	\begin{table}[!t]
		\centering
		\begin{tabular}{ccc}
			\hline
			Variable&Type&Dimension\\
			\hline
			x&list&150. Each element is a $10\times10\times 10$ array.\\
			y&vector&150\\
			z&matrix&$150\times 2$\\
			vec\_x&matrix&$1000\times 150$\\
			testx&list&1000. Each element is a $10\times10\times 10$ array.\\
			testy&vector&1000\\
			testz&matrix&$150\times 2$\\
			vec\_testx&matrix&$1000\times 1000$\\
			\hline
		\end{tabular}
		\caption{Data set dat.ten.}
		\label{tab:datten}
	\end{table}
	
	Function \fct{catch} fits a CATCH model on the input tensor data. Covariates are optional for the function and the function will fit a TDA model when there is no covariate. Function \fct{catch} has already integrated the adjustment step and model fitting step, hence it will automatically adjust for covariates when covariates exist. If one prefers to seperate the adjustment step, he/she can call \fct{adjten} function to make adjustments and then supply the adjusted predictors into \fct{catch}, which we will discuss in supportive functions.
	
	Similar to the two functions above, function \fct{catch} can generate a solution path on default or user specified potential values of the parameter. It will also perform prediction when testing data is specified. To make predictions on CATCH model, user can directly apply \fct{catch} function or separating adjustment and model fitting step and then call the \fct{predict} function. The following example shows how to fit the model and make prediction when covariates exist. As mentioned above, functions \fct{dsda} and \fct{msda} shares the same arguments name for covariates.

\begin{CodeChunk}
\begin{CodeInput}
R> obj <- catch(dat.ten$x, dat.ten$z, dat.ten$y, dat.ten$testx, dat.ten$testz)
R> pred <- obj$pred
R> err <- apply(pred, 2, function(x){mean(x!=dat.ten$testy)})
R> min(err)
\end{CodeInput}
\begin{CodeOutput}
[1] 0.167
\end{CodeOutput}
\begin{CodeInput}
R> obj$lambda[which.min(err)]
\end{CodeInput}
\begin{CodeOutput}
[1] 0.4270712
\end{CodeOutput}
\end{CodeChunk}

	An example of applying CATCH to fit model and perform prediction on CSA data is as follows. \fct{catch} function takes list of multi-dimensional array as input. In the dataset, \code{x} is a list of length 148, where each element is a matrix of dimension $36\times 3$; \code{y} is a vector whose value ranges between 1 and 21. We use default parameter sequence of length 100 and the prediction for each value of parameter is generated.

\begin{CodeChunk}
\begin{CodeInput}
R> data(csa)
R> x <- csa$IDLH
R> y <- csa$y
R> teindex <- seq(1,147,7)
R> obj <- catch(x[-teindex, ], y=y[-teindex], testx=x[teindex, ], nlambda=10)
R> err <- apply(obj$pred, 2, function(x){mean(x!=y[teindex])})
R> print(err)
\end{CodeInput}
\begin{CodeOutput}
[1] 0.9523819 0.1904762 0.0952381 0.0000000 0.0952381 0.0000000 0.0000000
  0.0000000 0.0000000 0.0000000
\end{CodeOutput}
\end{CodeChunk}

	\subsection{Other functions}\label{Sec:otherfun}
	
	\noindent{\bf Two special cases}

	We provide two more functions for two common problems in practice, binary classification and matrix classification, respectively. First,	the function \fct{catch\_matrix} fits CATCH model on matrix data (2-way tensor), which is a special case of \fct{catch}. The usage of \fct{catch\_matrix} is exactly the same as that of \fct{catch}, with the only exception that the predictor has to be a matrix instead of higher-order tensor.

	Second, our package includes the function \fct{dsda.all} that integrates cross validation, model fitting and prediction. It requires the input of the training set and testing set. Then the optimal tuning parameter is chosen by cross validation on the training set, and the corresponding testing error is reported. See the following example. 
	
		\begin{CodeInput}
R> obj <- dsda.all(dat.vec$x, dat.vec$y, dat.vec$testx, dat.vec$testy, nfolds = 10)	
R> print(obj$err)	
\end{CodeInput}
\begin{CodeOutput}
[1] 0.116
\end{CodeOutput}

	\noindent{\bf Supportive functions}

	The package provides functions \fct{cv.dsda}, \fct{cv.msda}, \fct{cv.SeSDA} and \fct{cv.catch} to perform cross validation. For all of these functions, user can give a sequence of potential values to tune parameter. Otherwise, the function will first fit a model on the entire data set and then perform cross validation on the automatically generated $\lambda$s from the entire data set. Similar as \fct{msda}, user can specify which model to use in \fct{cv.msda} or let the function determine by input data.
	
	Users can also specify the number of folds by the argument \code{nfolds}. Another argument \code{lambda.opt} has two options \code{"min"} and \code{"max"}. When multiple $\lambda$s lead to same error rate, \code{"min"} will return the smallest tuning parameter with the lowest error rate while \code{"max"} will return the largest one. We take \fct{cv.dsda} and \fct{cv.catch} as two examples. 
	
\begin{CodeChunk}
\begin{CodeInput}
R> obj.dsda <- cv.dsda(dat.vec$x, dat.vec$y, nfolds = 10)
R> obj.catch <- cv.catch(dat.ten$x, dat.ten$z, dat.ten$y, lambda.opt="min")
\end{CodeInput}
\end{CodeChunk}
	
	Function \fct{adjten} and \fct{adjvec} implement the adjustment step for tensor and vector data, respectively. It takes training tensor/vector, covariate and response as input, and outputs the adjusted tensor/vector and adjustment coefficients $\bolalpha$. The adjustement step has already been incorporated into the modeling fitting functions \fct{dsda}, \fct{msda} and \fct{catch}. When user input covariates along with tensor/vector, the model fitting functions will automatically make the adjustment. But if user do not want to use the automatic prediction in model fitting function and prefer to predict via \fct{predict}, user need to first make the adjustment to obtain adjustment coefficient \code{gamma}, and pass it into function \fct{predict}. Notice that making adjustment and fitting a model on the adjusted tensor and response without covariate is equivalent as fitting a model by inputting the original tensor, covariate and response labels. Examples of two approaches are given as follows.

\begin{CodeChunk}
\begin{CodeInput}
R> obj <- catch(dat.ten$x, dat.ten$z, dat.ten$y, dat.ten$testx, dat.ten$testz)
\end{CodeInput}
\end{CodeChunk}

\begin{CodeChunk}
\begin{CodeInput}
R> obj.adj <- adjten(dat.ten$x, dat.ten$z, dat.ten$y, dat.ten$testx,
    dat.ten$testz)
R> obj.fit <- catch(dat.ten$x, dat.ten$z, dat.ten$y)
R> pred <- predict(obj.fit, obj.adj$testxres, dat.ten$z, dat.ten$testz, 
    obj.adj$gamma)
\end{CodeInput}
\end{CodeChunk}

	There are three prediction functions corresponding to \fct{dsda}, \fct{msda} and \fct{catch}, respectively. All of them can be directly called by \fct{predict} and the function will recognize which function to use based on the input fitted model object. When covariate exists, user needs to pass the adjustment coefficient obtained from function \fct{adjten}, the fitted model and testing data altogether to make predictions. Therefore, we encourage user to direct use model fitting functions \fct{msda} and \fct{catch} to fit model and predict categorical responses.

	\section{Real Data Example}\label{sec: data}
	
	In this section, we will show the performance of the models by a real data set. 
	We considered the attention deficit hyperactivity disorder (ADHD) data set. The dataset is available on NITRC (\url{http://fcon\_1000.projects.nitrc.org/indi/adhd200}) \citep{ADHD}. It contains three parts of information: s-MRI data which is a 3-D tensor, covariate information including age, gender and handedness which is a vector, and response label. Among all 930 individuals, there are four types of categorical labels: Typically Developing Childrem (TDC), ADHD Combined, ADHD Hyperactive and ADHD Inattentive.
	
	We downsize the tensor to dimension $24\times 27\times 24$ and consider two classification scenarios. One is to combine ADHD Hyperactive with the ADHD Combined since there are only 13 subjects in class ADHD Hyperactive. This results in a multi-class problem with three classes. The second one is to further combine TDC and ADHD Inattentive since none of these two categories have hyperactivity symptoms. This give us a binary problem. We split the dataset into a training set and testing set by ratio $8:2$.
	
	Given the tensor structure and existence of covariates, the most suitable approach is to apply CATCH on that. We also vectorize the tensor into vectors and stack covariates along with the long vector to apply vector methods. For binary case, DSDA is applied. For multi-class case, MSDA model with \code{multi.modified} is applied since the dimension is too large to employ \code{multi.original}. We also compared with SOS \citep{Clemmensen} by its own package \pkg{sparseLDA}, $\ell_1$-GLM \citep{glmnet} by package \pkg{glmnet} and $\ell_1$-SVM \citep{cortes1995support,SVM,bradley1998feature,fung2004feature,becker2009penalizedsvm} by package \pkg{penalizedSVM}.

	\begin{table}[!t]
		\centering
		\begin{tabular}{ccccccc}
			\hline
			Method&\multicolumn{3}{c}{Binary}&\multicolumn{3}{c}{Multi-class}\\
			Error rate (\%) / Time (seconds)&Mean&SE&Time&Mean&SE&Time\\
			\hline
			DSDA/multi.modified&23.58&0.23&36.30&35.85&0.24&88.8\\
			SeSDA&23.68&0.24&731.28&NA&NA&NA\\
			CATCH&22.79&0.24&78.6&35.22&0.25&101.4\\
	 		$\ell_1$-GLM&23.99&0.16&36.23&35.66&0.21&134.4\\
			SOS&23.87&0.26&100.8&37.07&0.29&1114.8\\
			$\ell_1$-SVM&27.54&0.31&2835&41.28&0.32&14596\\
			\hline
		\end{tabular}
		\caption{ADHD classification. Average error rates based on 100 replicates and running time of 20 replicates are reported.}
		\label{tab: adhd}
	\end{table}
	
	For each replicate, we perform cross validation on training data and record the classification error on testing data. The entire process was repeated for 100 times and we report the mean and standard error of the error rates. The performance is shown in Table~\ref{tab: adhd}.

	\section{Discussion}
	
	Package \pkg{TULIP} provides a toolbox to fit various sparse discriminant analysis models, including parametric models DSDA, ROAD, SOS for binary vector data, semiparametric model SeSDA for binary vector data, MSDA for multiclass vector data, and CATCH for multiclass tensor data. As a comprehensive toolbox, the package provides prediction and cross validation functions as well. 
	
	Meanwhile, the package propose an approach to handle cases when both predictor and covariates are supplied. The predictor can be vector and tensor, while the covariates are usually low-dimensional vectors. Covariates may have an effect on both response and the predictor. Therefore making adjustment and excluding the influence of covariates from predictor is important. The package includes functions to make the adjustments and can be called easily by supplying covariates in the model fitting function.
	
\appendix
\section*{Appendices}\label{sec:appendix}
\addcontentsline{toc}{section}{Appendices}
\renewcommand{\thesubsection}{\Alph{subsection}}

\subsection{Tensor Notation} \label{Append.A}

	 On each dimension, which is named mode, a tensor is composed by vectors of length $(p_k\times 1)$ called mode-$k$ fiber, defined as $A_{i_1\cdots i_{k-1}I_ki_{k+1}\cdots i_M}$, $I_k=1,\dots p_k$. Stacking the mode-$1$ fiber by row gives the vectorization of a tensor $\vecc(\mbA)$, which is a $(\prod_{m}p_m\times1)$ column vector. If we unfold the tensor along the $k$-th mode, we obtain a matrix $\mbA_{(k)}\in\mathbb{R}^{p_k\times \prod_{l\ne k}p_l}$.

	Denote the \emph{mode-$k$ product} of a tensor $\mbA$ and a matrix $\bolalpha\in\mbbR^{d\times p_{k}}$ by $\mbA\times_{k}\bolalpha\in\mbbR$, which results in a tensor of dimension $p_{1}\times\cdots\times p_{k-1}\times d\times p_{k+1}\times\cdots\times p_M$. Each element of the product is the product of a mode-$k$ fiber of $\mbA$ and a row vector of $\bolalpha$. In particular, the \emph{mode-$k$ vector product} of a tensor $\mbA$ and a vector $\mbc\in\mbbR^{p_k}$ is a $(M-1)$-way tensor as a special case when $d=1$.
	The \emph{Tucker decomposition} of a tensor is defined as $\mbA = \mbC\times_{1}\mbG_1\times_{2}\cdots\times_{M}\mbG_M$, in short of $\llbracket\mbC;\mbG_1,\dots,\mbG_m\rrbracket$. In particular, the vectorization of tucker decomposition has the fact that $\vecc(\llbracket\mbC;\mbG_1,\dots,\mbG_M\rrbracket)=\left(\mbG_M\otimes\cdots\otimes\mbG_1\right)\vecc(\mbC)$, where $\otimes$ denotes Kronecker product.	
	If $\mbX=\bolmu+\llbracket \bZ; \bolSigma_1^{1/2},\ldots,\bolSigma_M^{1/2}\rrbracket$, where $\mbZ\in\mathbb{R}^{p_1\times\cdots\times p_M}$ and all elements of $\mbZ$ independently follow the univariate standard normal distribution, we say $\mbX$ follows a tensor normal distribution $\mbX\sim TN(\bolmu, \bolSigma_1,\ldots,\bolSigma_{M})$. The dependence structure on the $j$-th mode is measured by $\bolSigma_j>0$. Hence, $\vecc(\mbX)=\vecc(\bolmu) + \bolSigma^{1/2}\vecc(\mbZ)$, where $\bolSigma=\bolSigma_M\otimes\cdots\otimes\bolSigma_1$.

\subsection{Simulation code}

Data sets dat.vec and dat.ten are used to illustrate usage of the functions. Detailed model settings are described in Section~\ref{sec: usage}. Here are the code to simulate the two data sets.

Code to simulate data set dat.vec:

\begin{CodeChunk}
\begin{CodeInput}
R> set.seed(123456)
R> sigma <- matrix(0.3, 500, 500)
R> diag(sigma) <- 1
R> dsigma <- t(chol(sigma))
R> #define beta and mean
R> beta <- matrix(0, nrow = 500, ncol = 1)
R> beta[1:10,1] <- 0.5
R> M <- matrix(0, nrow = 2, ncol = 500)
R> M[2,] <- sigma%*%beta
R> y <- c(rep(1, 75), rep(2, 75))
R> #generate test data
R> telabel <- ceiling(runif(1000)*2)
R> x <- matrix(rnorm(150*500),ncol = 500)%*%t(dsigma)
R> x[y==2, ] <- x[y==2, ] + M[2,]
R> testx <- matrix(rnorm(1000*500), ncol = 500) %*% t(dsigma)
R> testx[telabel==2, ] <- testx[telabel==2, ] + M[2, ]
R> dat.vec <- list(x = x, y = y, testx = testx, testy = telabel)
\end{CodeInput}
\end{CodeChunk}	

Code to simulate data set dat.ten:

\begin{CodeChunk}
\begin{CodeInput}	
R> set.seed(123456)
R> sigma <- array(list(), 3) #define covariance matrices
R> dsigma <- array(list(), 3)
R> for (i in 1:3){
R>    sigma[[i]] <- diag(10)
R>    dsigma[[i]] <- t(chol(sigma[[i]]))
R> }
R> B2 <- array(0, dim=c(10,10,10)) #define B and mean
R> B2[1:2, 1:2, 1:2] <- 0.8
R> M <- array(list(), 2)
R> M[[1]] <- array(0, dim=c(10,10,10))
R> M[[2]] <- atrans(B2, sigma)
R> y <- c(rep(1,75), rep(2,75))
R> coef <- array(0, dim=c(10,10,10,2)) #define alpha
R> coef[1:5, 1:5, 1:5, 1] <- 1
R> telabel <- ceiling(runif(1000)*2)  
R> z <- matrix(rnorm(2*150), nrow=150, ncol=2) #generate covariates
R> z[y==2,] <- z[y==2,] + 0.3
R> testz <- matrix(rnorm(2*1000), nrow=1000, ncol=2)
R> testz[telabel==2, ] <- testz[telabel==2, ] + 0.3
R> vec_x <- matrix(rnorm(1000*150), ncol=150) #generate tensor
R> x <- array(list(),150)
R> for (i in 1:150){
R>    x[[i]] <- array(vec_x[,i], c(10,10,10)) + amprod(coef, t(z[i,]), 4)[,,,1]
R>    x[[i]] <- M[[y[i]]] + atrans(x[[i]], dsigma)
R> }
R> vec_testx <- matrix(rnorm(1000*1000), ncol=1000)
R> testx <- array(list(), 1000)
R> for (i in 1:1000){
R>    testx[[i]] <- array(vec_testx[,i], c(10,10,10)) + amprod(coef, t(testz[i,]),
	4)[,,,1]
R>    testx[[i]] <- M[[telabel[i]]] + atrans(testx[[i]], dsigma)
R> }
R> dat.ten <- list(x=x, z=z, testx=testx, testz=testz, vec_x=t(vec_x), 			   
vec_testx=t(vec_testx), y=y, testy=telabel)
\end{CodeInput}
\end{CodeChunk}	

\subsection{Estimation of covariance matrices in the TDA/CATCH model}\label{Appendix.C}

Denote the sample mean of Class $k$ by $\overline{\mbX}_k$. We first center $\mbX_i$ within class to obtain the residuals:
	$$
	\widehat{\mbE}_{i}=\mbX_{i}-\hatbolmu_k=\mbX_{i}-\overline{\mbX}_k.
	$$
Further unfold $\widehat{\mbE}_{i}$ along the $j$-th mode to obtain $\mbW_{i(j)}$ and find $\widetilde{\mbS}_j=(n\prod_{l\neq j}^{M}p_l)^{-1}\sum_{i=1}^{n}\mbW_{i(j)}(\mbW_{i(j)})^T$. Then our estimator for $\bolSigma_j$ is defined as
		\bea\label{sigma.hat}
	\hatbolSigma_{j}=\widetilde s_{j,11}^{-1}\widetilde{\mbS}_j \mbox{ for $j=1,\ldots,M-1$}; \hatbolSigma_{M}=\dfrac{\widehat{\Var}(X_{1 \cdots 1})}{\prod_{j=1}^M \widetilde s_{j,11}}\widetilde{\mbS}_M.
	\eea

\subsection{Definition of F-test statistic}
The F-test statistic used to preprocess GDS1615 data is defined as
\begin{equation}
f_j=\frac{\sum_{k=1}^Kn_k(\hatbolmu_{kj}-\widehat{\bar{\bolmu}}_j)^2/(K-1)}{\sum_{i=1}^n(\mbX_j^i-\hatbolmu_{Y^i,j})^2/(n-K)},
\end{equation}
where $\mbX_j^i$ is the $j$-th variable of $i$-th observation and $\widehat{\bar{\bolmu}}$ is the grand mean. 

\bibliography{draft}

\end{document}